\documentclass[twocolumn,aps,pre,superscriptaddress,showpacs]{revtex4}
\usepackage{graphicx}
\usepackage{color} 
\input{epsf}

\begin{document}

\title{Synchronization theory of microwave induced zero-resistance states}

\author{O.V.Zhirov}
\affiliation{\mbox{Budker Institute of Nuclear Physics, 
630090 Novosibirsk, Russia}}
\author{A.D.Chepelianskii}
\affiliation{\mbox{Cavendish Laboratory, Department of Physics, 
University of Cambridge, CB3 0HE, United Kingdom}}
\author{D.L.Shepelyansky}
%\homepage[]{http://www.quantware.ups-tlse.fr}
\affiliation{\mbox{Laboratoire de Physique Th\'eorique du CNRS, IRSAMC, 
Universit\'e de Toulouse, UPS, 31062 Toulouse, France}}

%\date{\today}
\date{February 12, 2013}

%\pacs{PACS numbers: 73.40.-c,05.45.-a,72.20.My}
%\PACS{
%{73.40.-c}{Electronic transport in interface structures}
%{72.20.My}{Galvanomagnetic and other magnetotransport effects} 
%{05.45.-a}{Nonlinear dynamics and chaos}
%{72.40.+w}{Photoconduction and photovoltaic effects}
%\and
%{73.63.-b}{Electronic transport in nanoscale materials and structures} 
%\and
%{05.45.Ac}{Low-dimensional chaos}
%}
%05.60.-k Transport processes 
%47.61.-k Micro- and nano- scale flow phenomena 
%81.07.-b Nanoscale materials and structures: 
%         fabrication and characterization in materials science)


\begin{abstract}
We develop the synchronization theory of
microwave induced zero-resistance states (ZRS)
for two-dimensional electron gas in a magnetic field.
In this theory the dissipative effects lead to synchronization of
cyclotron phase with driving microwave phase at certain
resonant ratios between microwave and cyclotron frequencies.  
This synchronization produces stabilization of 
electron transport along edge channels and 
at the same time it gives suppression of 
dissipative scattering on local impurities and dissipative conductivity 
in the bulk,
thus creating the ZRS phases at that frequency ratios.
The electron dynamics along  edge and around circular disk impurity
is well described by the Chirikov standard map.
The theoretical analysis is based on 
extensive numerical simulations of classical electron transport
in a strongly nonlinear regime. 
We also discuss the value of activation energy obtained in our model 
and the experimental signatures that could establish the synchronization 
origin of ZRS. 
\end{abstract}

\maketitle

\section{Introduction}

The experiments on resistivity of high mobility two-dimensional electron
gas (2DEG) in presence of a relatively weak magnetic field
and microwave radiation led to a discovery
of striking Zero-Resistance States (ZRS) induced by a microwave field
by Mani {\it et al.} \cite{mani2002} and Zudov {\it et al.}
\cite{zudov2003}. 
Other experimental groups also found
the  microwave induced ZRS in various 2DEG samples
(see e.g. \cite{dorozhkin,smet,bykov}).
A similar behavior of resistivity
is also observed for electrons on a surface of
liquid helium in presence of magnetic and microwave fields
\cite{konstantinov,adc}. These experimental results
obtained with different systems stress the generic nature of ZRS.
Various theoretical explications for this striking  phenomenon
have been proposed during the decade after the first experiments 
\cite{mani2002,zudov2003}. An overview of experimental and theoretical results
is give in the recent review \cite{dmitrievrmp}.

In our opinion the most intriguing feature of ZRS
is their almost periodic structure
as a function of the ratio $j=\omega/\omega_c$ between the microwave
frequency $\omega$ and cyclotron frequency $\omega_c=eB/mc$.
(in the following we are using units with electron charge $e$ 
and mass $m$ equal to unity). Indeed, a Hamiltonian 
of electron in a magnetic field
is equivalent to an oscillator, it has a magneto-plasmon resonance at
$j=1$ but in a linear oscillator there are no matrix elements at
$j=2,3,...$ and  hence a relatively weak microwave field
is not expected to affect electron dynamics and 
resistivity properties of transport. Of course, one can
argue that impurities can generate harmonics being resonant at high $j>1$
but ZRS is observed only in high mobility samples
and thus the density of impurities is expected to be rather low. 
It is also important to note that ZRS appears 
at high Landau levels $\nu \sim 50$ so that a semiclassical
analysis of the phenomenon seems to be rather relevant.

In this work we develop the theoretical approach proposed in
\cite{adcdls}. This approach argues that impurities produce only smooth
potential variations inside a bulk of a sample so that
ZRS at high $j$ appear from the orbits moving 
along sharp sample boundaries. It is shown \cite{adcdls} that 
collisions with boundaries naturally generate high
harmonics and that a moderate microwave field
gives stabilization of edge channel transport of 
electrons in a vicinity of $j \approx j_r = 1+1/4, \; 2+1/4, \; 3+1/4 ...$
producing at these $j$ a resistance going to zero 
with  increasing microwave power. This theory is based
on classical dynamics of electrons along a sharp edge.
The treatment of relaxation processes is modeled
in a phenomenological way by a dissipative term in 
the Newton equations. Additional noise term in the dynamical equations
takes into account thermal fluctuations. The dissipation 
leads to synchronization of cyclotron phase with
a phase of microwave field producing stabilization of edge
transport along the edges in a vicinity of resonant
$j_r$ values. Thus, according to the edge stabilization theory \cite{adcdls}
the ZRS phase is related to a universal synchronization
phenomenon which is a well established concept in nonlinear sciences
\cite{pikovsky}.

While the description of edge transport stabilization \cite{adcdls}
captures a number of important features observed in ZRS experiments
it assumes that the contribution of bulk orbits in transport
is negligibly small. This assumption is justified for smooth potential
variations inside the bulk of a sample. However, a presence of isolated 
small scale scatterers inside the bulk combined with a
smooth potential component can significantly affect the 
transport properties of electrons (see e.g. \cite{alikbyk}).
Also the majority of theoretical explanations of 
ZRS phenomenon considers only a contribution of 
scattering in a bulk \cite{dmitrievrmp}.
Thus  it is necessary to analyze how a scattering on
a single impurity is affected by  a combined action of
magnetic and microwave fields. In this work we perform 
such an  analysis modeling impurity 
by a rigid circular disk of finite radius.
We show that the dynamics in a vicinity of disk
has significant similarities with dynamics of orbits 
along a sharp edge leading to appearance of ZRS type features
in a resistivity dependence on $j$.

The paper is composed as follows: in Section  II
we discuss the dynamics in edge vicinity,
in Section III we analyze scattering on a single disk,
in Section IV we study scattering on many
disks when their density is low, here we determine the resistivity
dependence on $j$ and other system parameters,
physical scales of ZRS effects are analyzed in Section V, 
effects of two microwave driving fields 
and other theory predictions are  considered in 
Section VI, discussion of the results
is given in Section VII.

We study various models which we list here for a reader convenience:
wall model described by the Newton equations (\ref{eq1}), (\ref{eq2}) with 
microwave field polarization perpendicular to the wall (model (W1)
equivalent to model (1) in \cite{adcdls});
the Chirikov standard map description (\ref{eq3}) 
of the wall model dynamics called model (W2)
(equivalent to model (2) in \cite{adcdls} at parameter $\rho=1$);
the single disk model with radial microwave field
called model (DR1); the  Chirikov standard map description (\ref{eq3}) 
of model (DR1) called model (DR2) (here $v_y \rightarrow v_r$ in (\ref{eq3}),
$\rho >1$); the model of a single disk in 
a linearly polarized microwave field and static electric field
called model (D1); the model of transport in a system with many disks called 
model (D2) which extends the model (D1);
extension of model (D2) with disk roughness and 
dissipation in space called model (D3);
the wall model (W2)  extended to 
two microwave fields is called model (W3).

\section{Dynamics in edge vicinity}

We remind first the approach developed in \cite{adcdls}.
Here, the classical electron dynamics is considered 
in a proximity of the Fermi surface and in a vicinity
of sample edge modeled as a specular wall. 
The motion is described by  Newton equations
\begin{equation}
d \mathbf{v}/d t = \mathbf{\omega_c} \times \mathbf{v} + 
\omega \vec{\epsilon} \cos \omega t -
\gamma(v) \mathbf{v} + I_{ec} + I_{s}
\label{eq1}
\end{equation}
where a dimensionless vector
$\vec{\epsilon} = e \mathbf{E}/ (m \omega v_F)$ 
describes microwave driving field $\mathbf{E}$. Here an electron
velocity $v$ is measured in units of Fermi velocity $v_F$
and $\gamma(v) = \gamma_0 (|\mathbf{v}|^2 - 1)$ describes a relaxation 
processes to the Fermi surface. We also use the
dimensionless amplitude of velocity oscillations 
induced by a microwave  field
$\epsilon=   e |\mathbf{E}|/ (m \omega v_F)$.
As in \cite{adcdls}, in the following we use units with $v_F=1$.
The last two terms $I_{ec}$ and $I_{s}$ in (\ref{eq1}) account for 
elastic collisions with the wall and small angle scattering.
Disorder scattering is modeled as random rotations of $\mathbf{v}$ by 
small angles  in the interval $\pm \alpha_i$ 
with Poissonian distribution over time interval
$\tau_i=1/\omega$. 
The amplitude of noise is assumed to be relatively small
so that the mean free path $\ell_e$ is much larger than 
the cyclotron radius $r_c=v_F/\omega_c$.
We note that the dissipative term
is also known as a Gaussian thermostat \cite{hoover}
or as a Landau-Stuart dissipation \cite{pikovsky}.
The dynamical evolution described by Eq. (\ref{eq1}) 
is simulated numerically using the Runge-Kutta method. 
Following \cite{adcdls} we call this system model (W1)
(equivalent to model (1) in \cite{adcdls}).

We note that for typical experimental ZRS parameters we have:
electron density $n_e=3.5 \cdot 10^{11} cm^{-2}$,
effective electron mass $m=0.065 m_e$,
microwave frequency $f=\omega/2\pi= 50 GHz$,
Fermi  energy $E_F=m v_F^2/2 =\pi n_e \hbar^2/m =0.01289 V$,
corresponding to $E_F/k_B= 149.5 Kelvin$, with
Fermi velocity 
$v_F =2.641 \cdot 10^7 cm/s$.
At such a frequency the cyclotron
resonance $\omega=\omega_c=eB/mc$ takes
place at $B=0.1161 Tesla$ with the
cyclotron radius $r_c=v_F/\omega_c = 0.8873 \mu m$.
At such a magnetic field we have
the energy spacing between Landau levels
$\hbar \omega=\hbar \omega_c = 0.2067  mV = 2.40 K \cdot k_B$
corresponding to a Landau level
$\nu = E_F/\hbar \omega_c \approx 62$.
For a microwave field strength $E=1 V/cm$
we have the parameter $\epsilon=eE/(m\omega v_F) = 0.003261$.
With these physical values of system parameters
we can always recover the physical quantities
from our dimensionless units
with $m=e=v_F=1$.

Examples of orbits running along the edge of specular wall are given 
in \cite{adcdls} (see Fig.1 there). A microwave field creates resonances
between the microwave frequency $\omega$ and a frequency of nonlinear 
oscillations of orbits colliding with the wall. Due to
a specular nature of this collisions the electron motion 
has high harmonics of cyclotron frequency that leads to 
appearance of resonances around $j =1,2,3,4...$
(there is an additional shift of approximate value $1/4$ to $j_r$ values 
due to a finite width of nonlinear resonance).

To characterize the dynamical motion it is useful 
to construct the Poincar\'e section
following the standard methods of 
nonlinear systems \cite{chirikov,lichtenberg}.
We consider the Hamiltonian case at $\gamma_0=0$ in absence of noise.
Also we choose a linear polarized microwave field
being perpendicular to the wall which is going along $x$-axis 
(same geometry as in \cite{adcdls}).
In this case the generalized momentum 
$p_x=v_x +By=y_c$ is an integral of motion since there are no 
potential forces acting on electron along the wall
(here we use the Landau gauge with a vector potential
$A_x=By$). The momentum $p_x$ determines a distance
$y_c$ between a cyclotron center and the wall,
which also remains constant in time. 
The Hamiltonian of the system has the form:
\begin{equation}
H= p_y^2/2 + (p_x-By)^2/2 + \epsilon \omega y \cos \omega t +V_w(y) \; ,
\label{eq2}
\end{equation}
where $V_w(y)$ is the wall potential being zero or infinity
for $y<0$ or $y \ge 0$. 
Thus,  we have here a so called case of one and half
degrees of freedom (due to periodic time dependence of Hamiltonian on time)
 and the Poincar\'e section
has continuous invariant curves in the integrable 
regions of phase space \cite{chirikov,lichtenberg}.

\begin{figure}[!htb]
\begin{center}
\includegraphics[width=0.9\columnwidth]{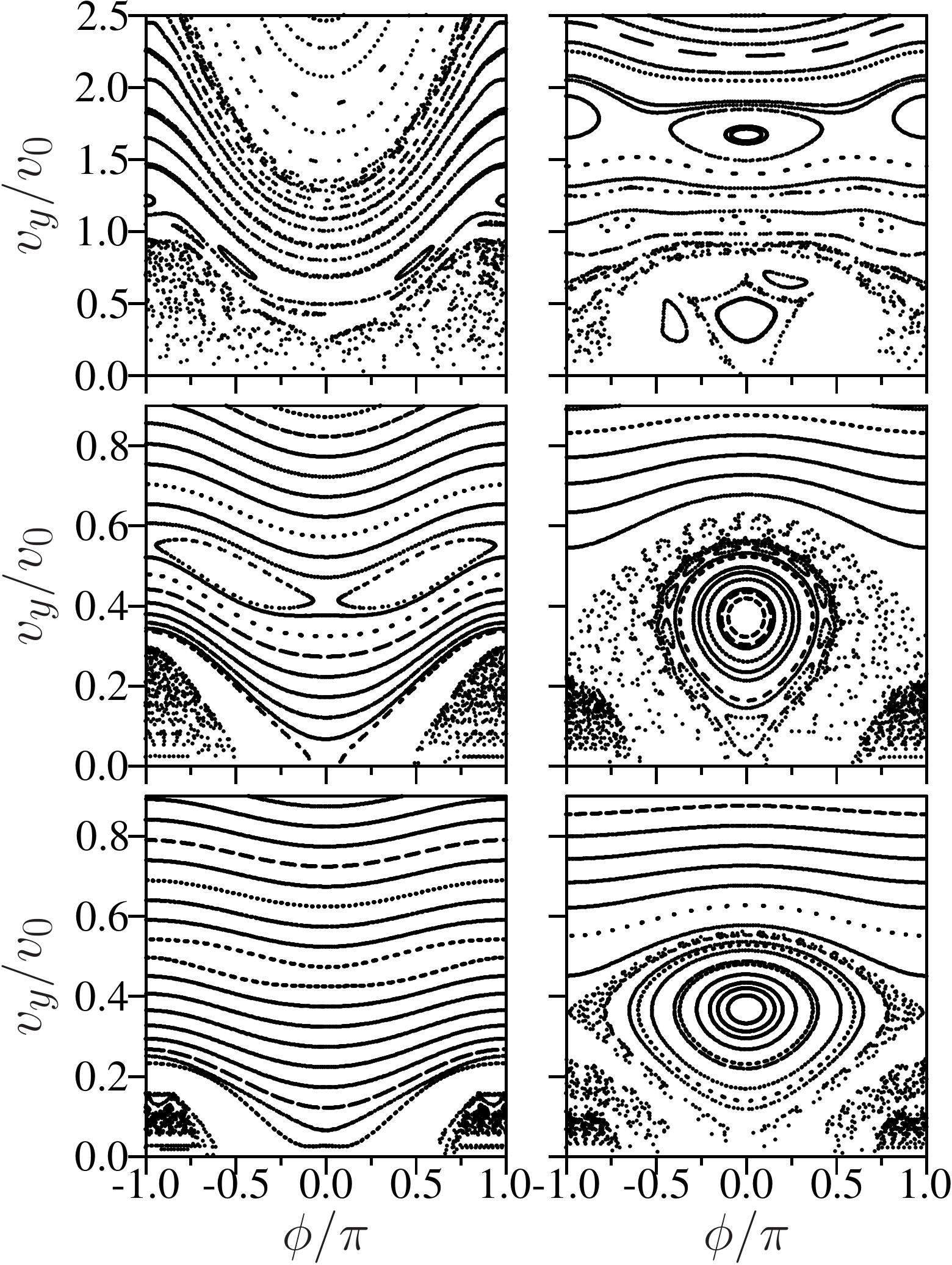}
\end{center}
\caption{ 
  Poincar\'e sections of Hamiltonian (\ref{eq2})
   for $j=7/4$ (left column) and $j=9/4$ (right column) 
  and different amplitudes of microwave field 
  $\epsilon=0.02, 0.04, 0.2$ (from bottom to top).
  Here, the integral $p_x/mv_F=1$, trajectories start from wall
   with fixed $v_x=p_x=v_0=1$. Data for model (W1) at $\gamma_0=0$,
   $\alpha_i=0$.
} 
\label{fig1}
\end{figure}

The Poincar\'e sections for (\ref{eq1}), (\ref{eq2}) 
at $j=7/4, 9/4$ and various amplitudes
of  microwave field $\epsilon$ are shown in Fig.~\ref{fig1}.
It shows a velocity $v_y$ at  moments of collision with the
wall at $y=0$ as a function of microwave phase $\phi = \omega t$ 
at these moments of time.
All orbits initially start at the wall edge $y=0$
with the initial velocity $v_x=v_0=p_x=y_c$. The value of $p_x=y_c$
is the integral of motion. However, the kinetic energy
of electron $E_k=(v_x^2+v_y^2)/2$ varies with time.
We see that at a small $\epsilon =0.02$ the main part of the
phase space is covered by invariant curves corresponding to
integrable dynamics. However, a presence of chaotic 
component with scattered points is also visible
in a vicinity of separatrix of resonances, especially
at large $\epsilon=0.2$. The points at $v_y $ close to zero 
correspond to orbits only slightly touching the wall,
while the orbits at $v_y/v_0 \gg 1$ have a large cyclotron radius
and collide with the wall almost perpendicularly.
There are also sliding orbits which have the center of cyclotron 
orbit inside the wall $(y_c>0)$ but we do not discuss they here.
Indeed, the orbits, which only slightly touch the wall
($y_c \approx - v_F/\omega_c$),  play the most important role 
for transport since the scattering angles in the bulk
are small for high mobility samples and
an exchange between bulk and edge goes
via such type of dominant orbits \cite{adcdls}.

We note that the section of Fig.~\ref{fig1} at
$j=9/4$, $\epsilon=0.02$ is in a good agreement with those shown
in Fig.1b of \cite{adcdls}. However, here we have single
invariant curves while in \cite{adcdls} the curves have a certain finite width.
This happens due to the fact that in \cite{adcdls} the Poincar\'e
section was done with trajectories having different 
values of the integral $p_x=y_c$ that
gave some  broadening of invariant curves.
For a fixed integral value we have no overlap
between invariant curves as it is well seen in Fig.~\ref{fig1} here.

The phase space in Fig.~\ref{fig1} has a characteristic
resonance at a certain $v_y/v_0$ value which 
position depends on $j$ \cite{adcdls}. An approximate description 
of the electron dynamics and phase space structure can be obtained
on a basis of the Chirikov standard map 
\cite{chirikov,lichtenberg},\cite{scholar}. 
In this description developed in \cite{adcdls} an electron velocity
has an oscillating  component  $\delta v_y = \epsilon \sin \omega t$
(assuming that $\omega > \omega_c$) and a collision with the wall
gives a change of modulus of $v_y$ by $2 \delta v_y$
(like a collision with a moving wall). 
For small collision angles the time between collisions is 
$\Delta t = 2 (\pi - v_y) / \omega_c$. 
Indeed, $2\pi/\omega_c$ is the cyclotron period.
However, the time between collisions is slightly smaller by
an amount $2v_y/\omega_c$: at $v_y \ll v_x \approx v_F$
an electron moves in an effective triangular well created by
the Lorentz force and like for a stone thrown against
a gravitational field this  gives the above reduction of 
$\Delta t$ (formally this expression for $\Delta t$
is valid for sliding orbits but for orbits slightly touching the wall
we have the same $\Delta t$ but with minus that gives the correction
 $-2v_y/\omega_c$). 
The same result can be obtained via semiclassical quantization
of edge states developed in \cite{avishai}.
It also can be found from a geometric overlap
between the wall and cyclotron circle.
This yields 
an approximate dynamics description 
in terms of the Chirikov standard map \cite{chirikov}:
\begin{equation}
{\bar v_y} = v_y + 2 \epsilon \sin \phi + I_{cc}, \;
{\bar \phi} = \phi + 2 (\pi - {\bar v_y/\rho}) \omega_ /\omega_c \; ,
\label{eq3}
\end{equation}
with the chaos parameter $K= 4 \epsilon \omega/(\rho \omega_c)$.
Usually we are in the integrable regime with $K <1$
due to small values of $\epsilon$ used in experiments. 
A developed chaos appears at $K>1$ \cite{chirikov,lichtenberg}.
Here bars mark the new values of variables going from 
one collision to a next one, $v_y$ 
is the velocity component perpendicular to the wall,
$\phi=\omega t$ is the microwave 
phase at the moment of collision.
Here we introduced a dimensionless parameter $\rho$
which is equal to $\rho=1$ for the case of the wall model
$W2$ considered here. However, we will show that
for the dynamics around disk with a radial field
in model (DR1) we have the same map (\ref{eq3})
with $\rho=1+r_c/r_d$. Due to that it is convenient
to write all formula with $\rho$.
We note that a similar map (\ref{eq3}) describes 
also a particle dynamics in a
one-dimensional triangular well and a monochromatic field
\cite{bubblon}.

The term $I_{cc} = - \gamma_c v_y + \alpha_n$ in (\ref{eq3}) describes 
dissipation and noise. The later  gives fluctuations of velocity $v_y$
at each  iteration ($-\alpha < \alpha_n <\alpha$; corresponding to
random rotation of velocity vector in (\ref{eq1})). 
Damping from electron-phonon and
electron-electron collisions contribute to $\gamma_c$.
The Poincar\'e sections of this map are in a good agreement
with those obtained from the Hamiltonian dynamics as it is seen
in Fig.~\ref{fig1} here and Fig.1 in \cite{adcdls}.
Following \cite{adcdls} we call this system model (W2)
(equivalent to model (2) in \cite{adcdls}).

A phase shift of $\phi$ by $2\pi$ does not affect the dynamics
and due to that the phase space structure 
changes periodically with integer values of $j$.
Indeed, the position of the main resonance
corresponds to a  change of phase by an integer number of $2\pi$
values ${\bar \phi} - \phi = 2\pi m = 2(\pi-v_y/\rho)\omega /\omega_c$
that gives the position of resonance at 
$v_{res}=v_y = \pi \rho (1 - m \omega_c/\omega) = \pi \rho \delta j/j$ 
where $m$ is the nearest integer  of $\omega/\omega_c$
and $\delta j$ is the fractional part of $j$.
Due to this relation we have the different resonance position
for $j=7/4$ and $9/4$ being in agreement with the data of 
Fig.~\ref{fig1} at small values of $\epsilon$ when nonlinear corrections
are small (we have here $\rho=1$). Thus at $j=9/4$ we have the resonance
position at $v_y= 0.1111 \pi \approx 0.35$ in agreement 
with Fig.~\ref{fig1} (right bottom panel).
For $j=2$ we have $v_y=0$ and at $j=7/4$ the resonance 
position moves to negative value $v_y = -0.45$. 
Thus, at $j=2; 7/4$ the resonance separatrix
easily moves particles out from the edge at $v_y<0$ where
they escape to the bulk due to noise.
In contrast at $j =9/4$ particles move along separatrix
closer to the edge being then captured inside the resonance
which gives synchronization of cyclotron phase with the microwave phase.
This mechanism stabilizes the transmission  along the edge.
\begin{figure}[!h]
\begin{center} 
\includegraphics[width=0.9\columnwidth]{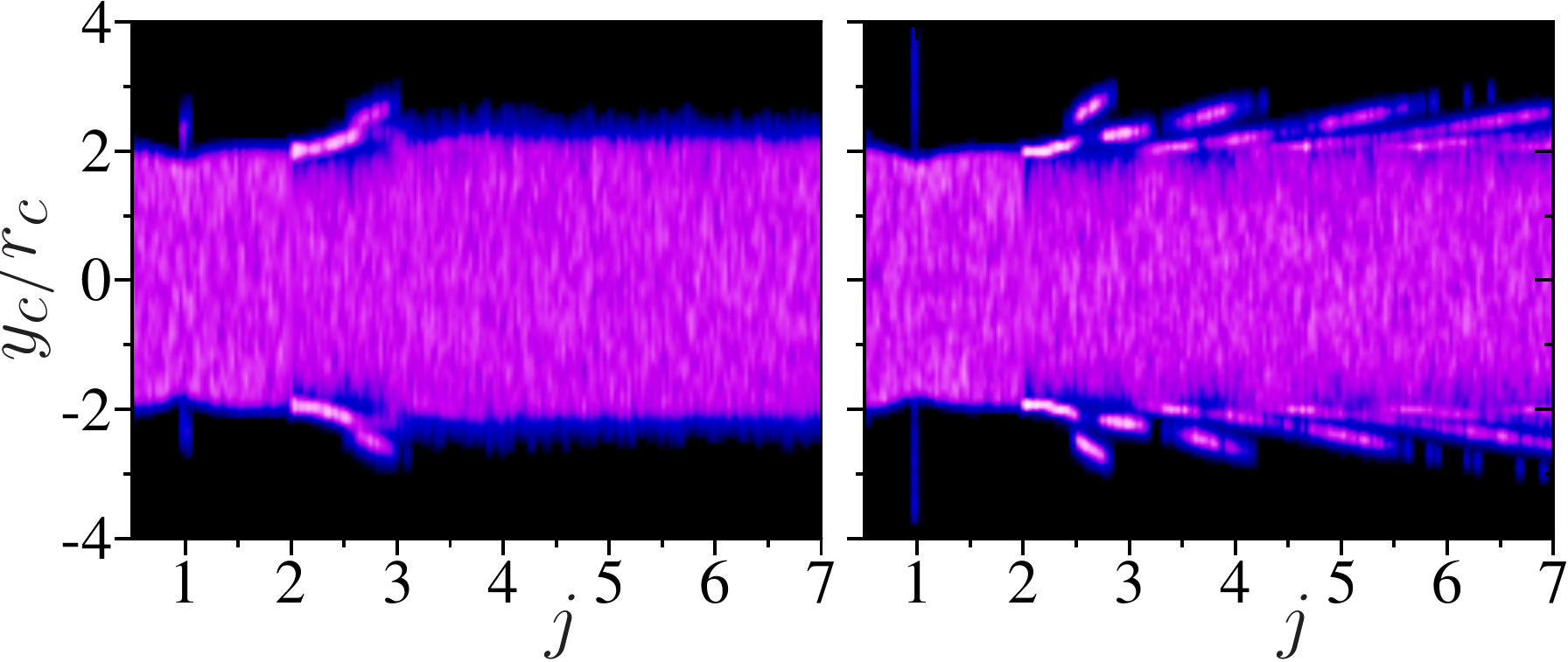}
\end{center}
\caption{ (Color online) Density distribution  $w$ of electrons
as a function of their dimensionless
cyclotron center position $y_c/r_c$
between two walls and the  frequency ratio
$j=\omega/\omega_c$. The distance between specular  walls is
$6 r_c$. The amplitude of microwave field is
$\epsilon =0.1$ with polarization parallel (left panel)
and perpendicular (right panel) to the walls
(see text for more details).
Here $\gamma_0/\omega=0.05$, $\alpha_i=0.01$, $\tau_i=1/\omega$,
The variation of $j=\omega/\omega_c$
is obtained by changing magnetic field ($\omega_c=B$)
keeping $\omega=const$; 100 electrons are simulated at each $j$
up to time $t_r=10^5/\omega$. 
Density is proportional to color changing from zero (black)
to maximal density (white). Data for model (W1).
} 
\label{fig2}
\end{figure}

In \cite{adcdls} it is shown that the orbits started 
in edge vicinity are strongly affected by a microwave field
that leads to ZRS type oscillations of transmission along the
edge and longitudinal resistivity $R_{xx}$.
The ZRS structure appears both in the frame of dynamics 
described by (\ref{eq1}) (model (W1))
and map description (\ref{eq3}) (model (W2)).
The physical mechanism is based on synchronization of 
a cyclotron phase with a phase of microwave driving
that leads to stabilization of electron transport
along the edge.  An extensive amount of numerical data
has been presented in \cite{adcdls}
and we think there is no need to add more.
Here, we simply want to illustrate that even those 
orbits which start in the bulk are affected by this 
synchronization effect. For that we take 
a band of two walls with a band width between them being
$\Delta y= L=6r_c$. Initially 100 trajectories are distributed randomly
in a bulk part between walls when a cyclotron radius is not
touching the walls $(-2r_c < y_c < 2r_c)$.
Their dynamics is followed during the run time
$t_r=10^5/\omega$ according to Eq.~(\ref{eq1})
and a density distribution $w(y_c)$ averaged 
in a time interval $5 \cdot 10^4 < \omega t < 10^5$ 
is obtained for a range of $0.5 \leq j \leq 7$
(261 values of $j$ are taken homogeneously in this
interval). The value of $t_r$ 
approximately corresponds to
a distance propagation along the wall
of $r_w \sim v_y t_r \sim 0.1 v_F t_r \sim 5 \cdot 10^3 v_F/\omega \sim
0.2 cm$ at typical values $v_F \sim 2\cdot 10^7 cm/s$, $\omega/2\pi =100 GHz$. 
This is comparable with a usual sample size used in 
experiments \cite{mani2002,zudov2003}. 
Similar values of $r_w$ were used in \cite{adcdls}.

\begin{figure}[!ht] 
\begin{center} 
\includegraphics[width=0.9\columnwidth]{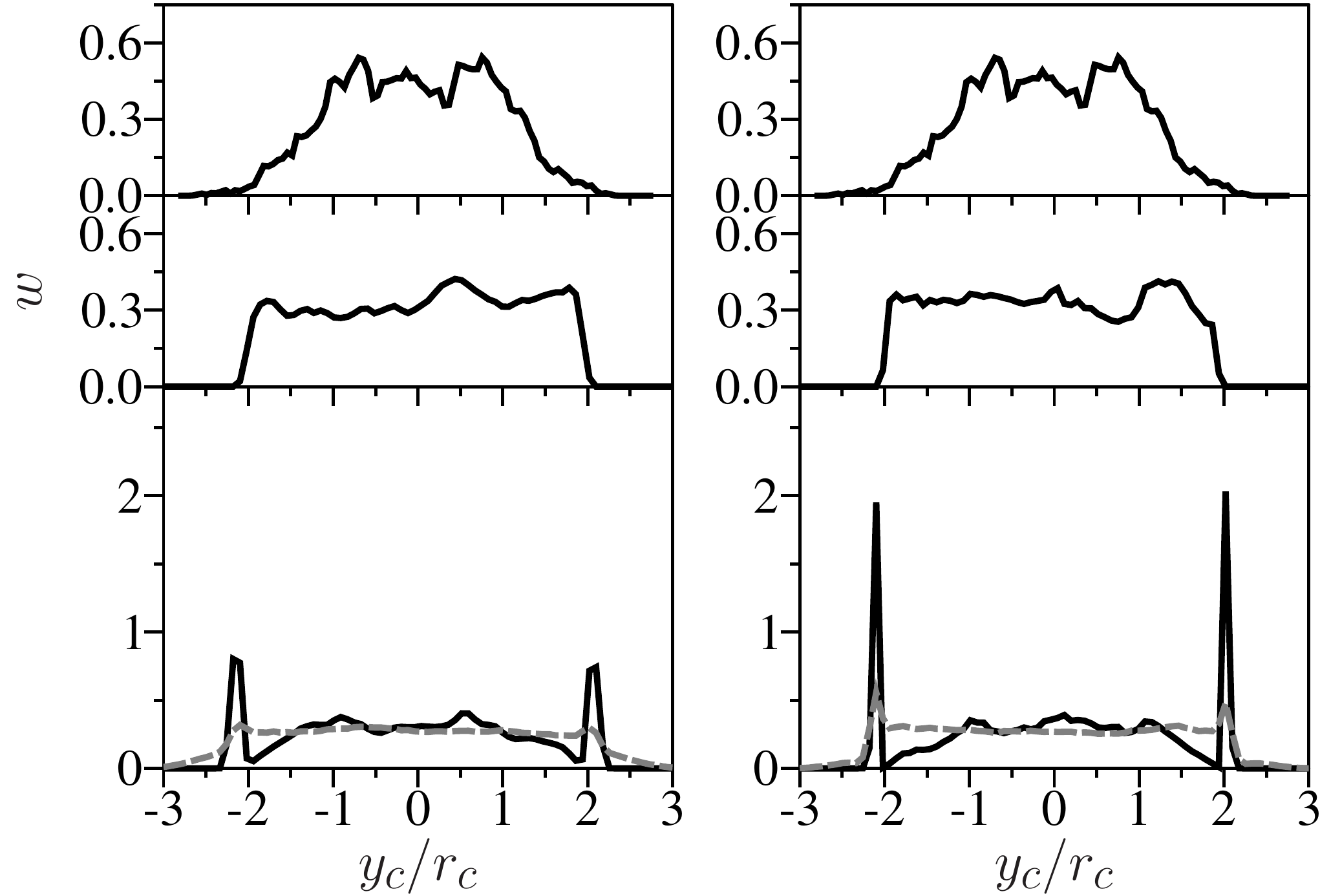}
\end{center}
\caption{Profile of density distribution $w(y_c)$
as a function of $y_c/r_c$ for microwave 
polarization parallel (left panels) and perpendicular (right panels)
to the walls. Here we have no microwave at top panels,
$\epsilon = 0.1$, $j=1.7$ at middle panels,
$\epsilon = 0.1$, $j=2.4$ at bottom panels.
In all panels we have noise amplitude $\alpha_i=0.01$
as in Fig.~\ref{fig2}, dashed curves in bottom
panels are obtained with $\alpha_i=0.05$.
Simulations are done with 500 trajectories,
other parameters are the same as in Fig.~\ref{fig2}.
Data for model (W1).
} 
\label{fig3}
\end{figure}

The dependence of density $w$ on
$y_c$ and $j$ are shown in Fig.~\ref{fig2}
for two polarizations of microwave field.
The data show that orbits from a bulk can be captured in edge vicinity
for a long time giving an increase of density in 
a vicinity of edge. This capture is significant
around resonance values $j \approx j_r$.
This is confirmed by a direct comparison of  
density profiles in Fig.~\ref{fig3}
at $j=1.7 \approx 2 -1/4$ and $j=2.4 \approx 2+1/4$.
In the later case we have a large density peak
due to trajectories trapped in a resonance 
(see Fig.~\ref{fig1}) where they are synchronized
with a microwave field.
An increase of noise amplitude 
$\alpha_i$ gives a significant reduction of the amplitude
of these resonant peaks (Fig.~\ref{fig3} bottom panels).
The increase of density is more pronounced for
polarization perpendicular to the wall
in agreement with data shown in 
Fig.2 of \cite{adcdls}. 

We also performed numerical simulations
using Eq.~(\ref{eq1}) with a smooth wall
modeled by a potential $V_w(y) = \kappa y^2/2$.
For large values $\kappa/\omega_c$ (e.g. $\kappa/\omega_c=10$)
we find the Poincar\'e sections to be rather similar to
those shown in Fig.~\ref{fig1} that gives 
a similar structure of electron density as in 
Figs.~\ref{fig2},\ref{fig3}. A finite wall rigidity 
can produce a certain shift of
optimal capture conditions appearing as a result of
additional correction
to a cyclotron period due to a part of orbit inside the wall. 

The data presented in this Section show that
electrons from the bulk part of the sample can be 
captured for a long time in edge vicinity thereby 
increasing the electron density near the edge.
This effect is very similar to the accumulation of electrons 
on the edges of the electron cloud under ZRS conditions that was reported 
for surface electrons on Helium in \cite{adc}. 
However we have to emphasize that the confinement potential for surface electrons 
is very different from the hard wall potential assumed in our simulations, 
as a consequence our results cannot be applied directly to this case.
It is possible that the formation of ballistic 
channels on the edge of the sample
combined with the redistribution of the electrons density
can effectively short the bulk contribution and 
induce directly a vanishing $R_{xx}$.
However, it is also important to understand 
how a scattering on impurities inside the bulk 
is affected by a microwave radiation.
We study this question in next Sections.

%\newpage

\section{Scattering on a single disk}

It should be noted that resistivity properties of
a regular lattice of disk antidots in 2DEG
had been studied experimentally \cite{weiss,kvon}
and theoretically \cite{geisel1,geisel2}.
But effects of microwave field were not considered
till present.

In our studies we model an impurity as a rigid disk of fixed radius
$r_d=v_F/\omega$ keeping $\omega = const$
and changing $\omega_c=B$.
In a magnetic field a cyclotron radius
moves in a free space only due to a static
{\it dc-}electric field $E_{dc}$.
We fix the direction of $E_{dc}$ along $x-$axis
and measure its strength by
a dimensionless parameter $\epsilon_s=E_{dc}/(\omega v_F)$.
Even in absence of a microwave field 
a motion in a vicinity of disk
in crossed static electric and magnetic fields
of moderate strength is not so simple. 
The studies presented in
\cite{berglund} and \cite{aliktrap}
show that dynamics in disk vicinity
is described by a symplectic disk map
which is rather similar to the map (\ref{eq3}).
It is characterized by a chaos parameter
$\epsilon_d = 2\pi v_d/(r_d \omega_c)$
where $v_d=v_F E_{dc}/B$ is the drift velocity;
$\epsilon_d$ gives an amplitude  of change of radial
velocity at collision. 
Orbits from a vicinity of disk can escape 
for $\epsilon_d >  0.45 $ \cite{berglund}.
\begin{figure}[!ht] 
\begin{center} 
\includegraphics[width=0.9\columnwidth]{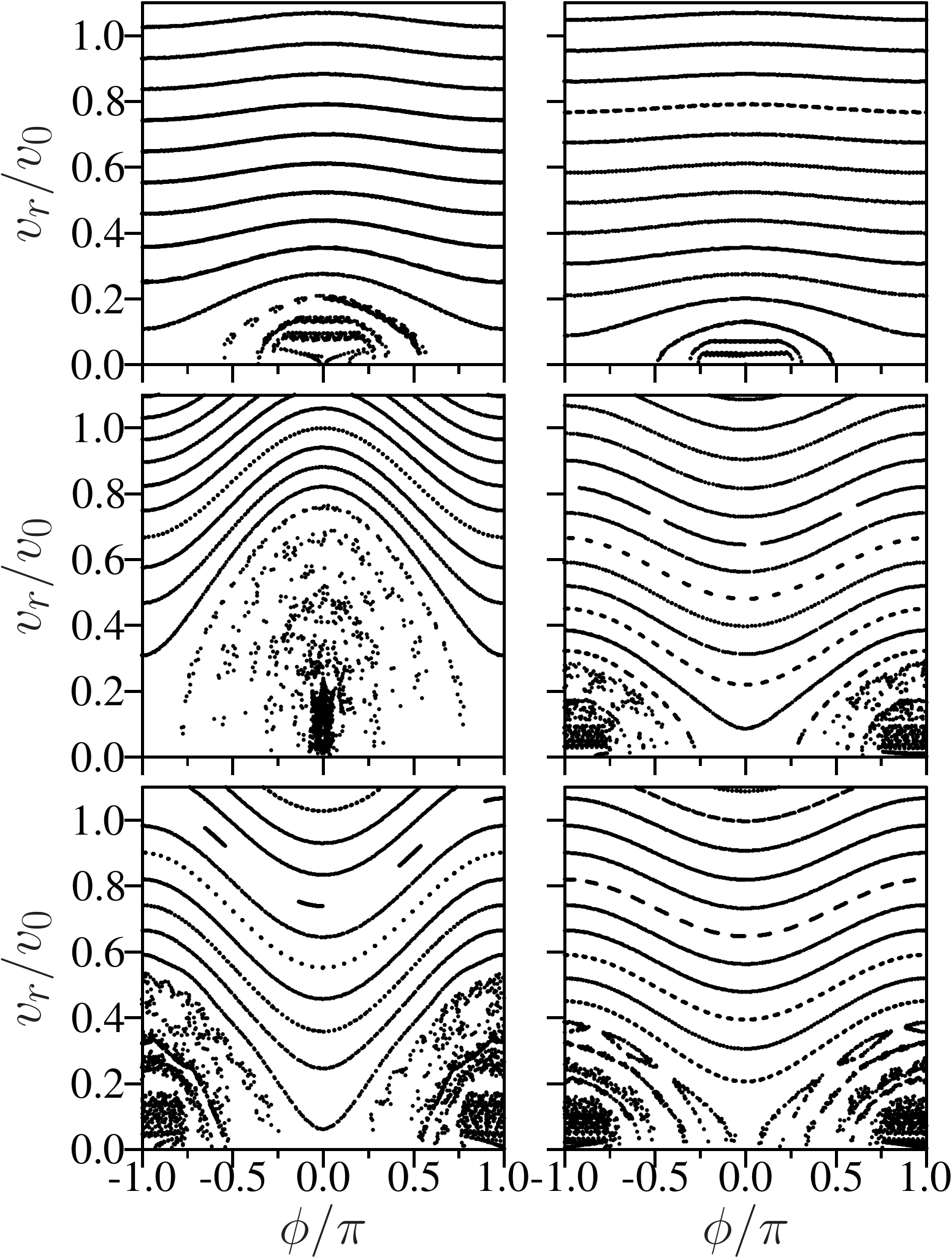}
\end{center}
\caption{ 
  Poincar\'e section for Hamiltonian dynamics in a disk vicinity in presence of
  radial microwave field. Left column panels: $j=7/4, 2, 9/4$ at
  $\epsilon=0.04$ (from top to bottom);
  right column panels:  $j=7/4, 13/4, 9/4$ at $\epsilon=0.02$.
  Here the integral of orbital momentum is $\ell_0/v_F r_d= v_0/v_F=1$,
  trajectories start from disk with fixed tangent velocity component $v_0=1$.
  Data for model (DR1).
} 
\label{fig4}
\end{figure}

We start our analysis from the construction of the Poincar\'e
section in presence of microwave field at zero static field.
To have a case with one and half degrees of freedom
we start from a model case when a microwave field
is directed only along radius from a disk center.
The dynamics is described by Eq.~(\ref{eq1})
with a dimensionless microwave amplitude $\epsilon$.
The dynamical evolution is obtained 
numerically by  the Runge-Kutta method.
At first we consider a case without dissipation and noise.
Due to radial force direction the orbital momentum
is an additional integral of motion (as $p_x=y_c$ 
for the wall case) and thus
we have  again 3/2 degrees of freedom.
We call this disk model with radial microwave field as model (DR1).

\begin{figure}[!ht] 
\begin{center} 
\includegraphics[width=0.9\columnwidth]{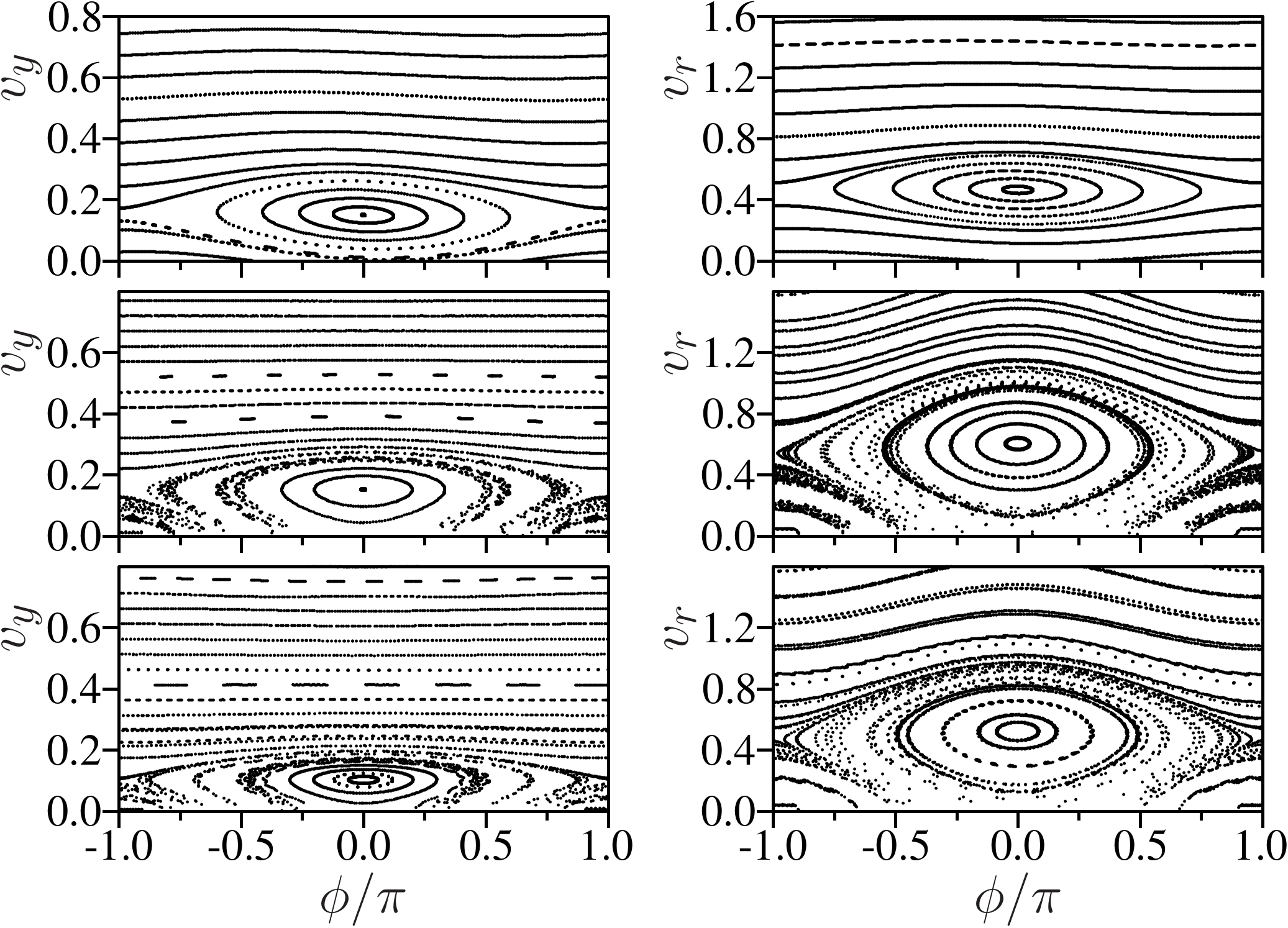}
\end{center}
\caption{Poincar\'e sections for wall model (W1) (left) and disk model 
  with radial electric field (DR1) (right) 
  at $\epsilon=0.01$ and $j=2.1$ (top, middle) and $j=3.1$ (bottom).
  Top panels are obtained from the Chirikov standard map (\ref{eq3})
  at $\rho=1$ (left top panel), corresponding to the wall model (W2),
   and at $\rho=1+j=3.1$ (right top panel),
   corresponding to the disk model (DR2).
  Middle and bottom panels are obtained from solution of Newton equations
  (\ref{eq1}) for wall (left) and disk (right).
  Other parameters are as in Fig.~\ref{fig1} and Fig.~\ref{fig4},
   $v_y$ and $v_r$ are expressed in units of $v_F$.
  There is no dissipation and no noise.
} 
\label{fig5}
\end{figure}

The Poincar\'e sections at the moments of collisions with disk
are shown in Fig.~\ref{fig4},~\ref{fig5} for model (DR1).
Here, $v_r$ is the radial component 
of electron velocity and $\phi$ is a microwave phase both taken
at the moment of collision with disk. We see that the phase space 
structure remains approximately the same when $j$ is increased by unity 
(compare $j=9/4, 13/4$ panels in Fig.~\ref{fig4}).
This happens for orbits only slightly touching the disk
(small $v_r$) since the microwave phase change
during a cyclotron period is shifted by an integer
amount of $2\pi$ (in a first approximation at $r_d \ll r_c$).
The similarity between the wall and disk cases is directly
seen from Fig.~\ref{fig5} as well as periodicity with
$j \rightarrow j+1$.

In fact in the case of disk with a radial field
the dynamics can be also described by the Chirikov standard
map (\ref{eq3}) where $v_y$ should be understood as a radial velocity
$v_r$ at the moment of collision. The second equation has 
the same form since the change of the phase between 
two collisions is given by the same equation but with the parameter
$\rho=1+r_c/r_d$. This expression for $\rho$
is obtained from the geometry of slightly
intersecting circles of radius $r_d$ for disk
and radius $r_c$ for cyclotron orbit
(the angle segment of cyclotron circle is $\Delta \varphi=2v_r/\rho$).
For $r_d \gg r_c$ this expression naturally reproduces
the wall case while at $r_d \ll r_c$
we have the correction term proportional to ${\bar v}_y$
going to zero that also well corresponds to the
geometry of two disks. After such modification of $\rho$
we find that the resonance positions $v_{res}=\pi \rho \delta j/j$
are proportional to $\rho$. Thus the  model (DR1)
reduced to the map description (\ref{eq3})
at $\rho>1$ is called model (DR2).

The expression for $v_{res}$ works rather well.
Indeed, for $j=2.1$ in Fig.~\ref{fig5} we obtain
$v_{res}=0.149$ for model (W2) and $0.463$ for model (DR2).
These values are in a good agreement with numerical values 
$v_{res} \approx 0.15$ for model (W2) and 
$v_{res} \approx 0.6$ for model (DR1). In the later case
the agreement is less accurate due to a larger size of nonlinear resonance.
The comparison of Poincar\'e sections given by
the Chirikov standard map (\ref{eq3})
and the dynamics from Newton equations, shown in Fig.~\ref{fig5},
confirms the validity of map description.

According to the well established results for the 
Chirikov standard map \cite{chirikov}
we find for models (W1), (W2) and (DR1), (DR2)
the width of separatrix $\delta v$ and
the corresponding resonance energy width
$E_r=(\delta v)^2/2$:
\begin{eqnarray}
\label{eq4}
E_r & = & 16 \epsilon \omega_c \rho E_F/\omega \; ; \; \rho=1+r_c/r_d \; ;\\
\nonumber
v_{res} & = & \pi \rho \delta j/j \; ; \; 
\delta v= 4  \sqrt{\epsilon \rho/j} \; ; \\
\nonumber
\delta j_\epsilon & = & \delta v j/(2\pi \rho) \; ; \; j=1+\omega/\omega_c \; ,
\end{eqnarray}
where $\delta j_\epsilon$ is the resonance shift produced 
by a resonance half width $\delta v/2 =v_{res}$.
This relation shows that for the disk case
this energy is increased by a factor $\rho$ compared to the wall case. 
In majority of our numerical simulations we have
$\rho=1+j$.

Thus a  radial field models (DR1), (DR2) 
represent a useful approximation to understand the 
properties of dynamics in a disk vicinity but a real 
situation corresponds to a linear microwave polarization
and the Poincar\'e section analysis 
should be modified to understand the dynamics in this case.

Due to that we start to analyze 
the scattering problem on a disk in presence 
of weak static field $\epsilon_s$
and microwave field $\epsilon$ using Eq.~(\ref{eq1}). 
For the scattering problem
we find more simple to have dissipation to work only at 
the time moments of electron collisions with disk: at such time moments
the radial component of electron velocity is reduced by a factor
$v_r \rightarrow v_r/(1+\gamma_d)$, the reduction is
done only if the kinetic energy of electron is larger
than the Fermi energy. Such a dissipation can be induced 
by phonon excitations inside the antidot disk.  
We fix geometry directing {\it dc-}field along $x-$axis and
microwave along $y-$axis. The noise is modeled in the same way as above in
Eq.~(\ref{eq1}). We call this system disk model (D1).
\begin{figure}[!ht] 
\begin{center} 
\includegraphics[width=0.9\columnwidth]{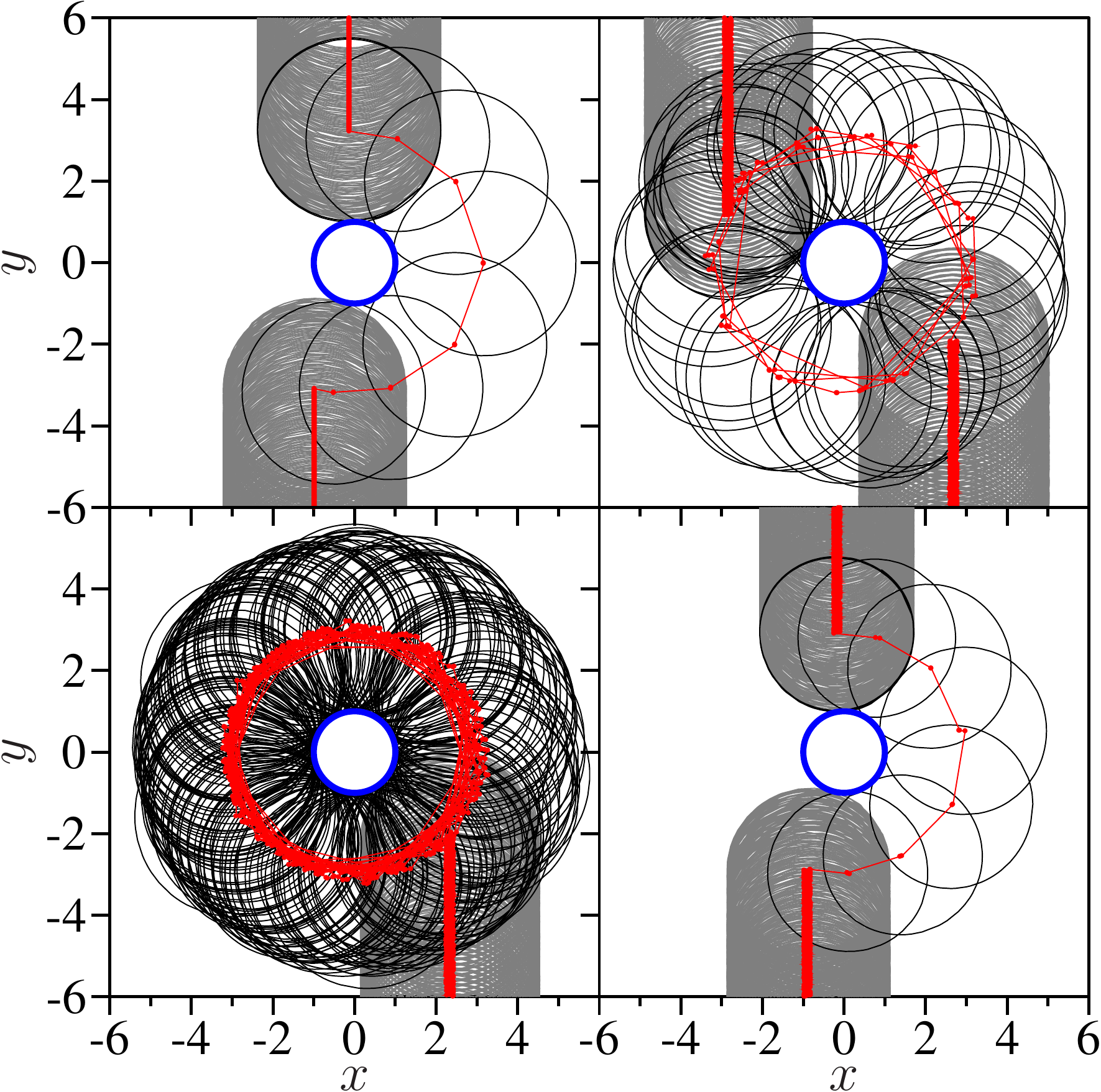}
\end{center}
\caption{(Color online) 
  Scattering of electron cyclotron trajectory on a disk scatterer
  (blue/black circle) in model (D1).
  \textit{Top left panel} shows the case in absence 
   of microwave, $\epsilon=0$, $j=9/4$. 
  \textit{Top right panel}: temporary captured path at 
  $\epsilon=0.04$, $j=9/4$. 
  \textit{Bottom left panel}: path captured forever at
   $\epsilon=0.04$, $j=9/4$. 
  \textit{Bottom right panel}: no capture at
   $\epsilon=0.04$, $j=2$. 
  The trajectory part colliding  with disk is shown by black curve, 
  its part before and after collisions is shown in gray. 
  The red (light gray)  points and curves show the 
  trajectory of cyclotron center. 
  Here the dissipation parameter is 
  $\gamma_d/\omega=0.01$;
  the static electric field is directed along 
  $x-$axis and $\epsilon_{s}=0.001$; microwave field is directed along
  $y-$axis. There is no noise here. Coordinates $x,y$ are expressed in
  units of $r_d$. Data for model (D1).
} 
\label{fig6}
\end{figure}

Examples of electron cyclotron trajectories scattering 
on disk are shown in Fig.~\ref{fig6}. In absence of microwave field
a trajectory escapes from disk rather rapidly. A similar 
situation appears at $j=2$ and microwave field with $\epsilon=0.04$.
In contrast for $j=9/4$ and $\epsilon=0.04$ a trajectory can be 
captured for a  long time or even forever depending 
on initial impact parameter.

\begin{figure}[!ht] 
\begin{center} 
\includegraphics[width=0.9\columnwidth]{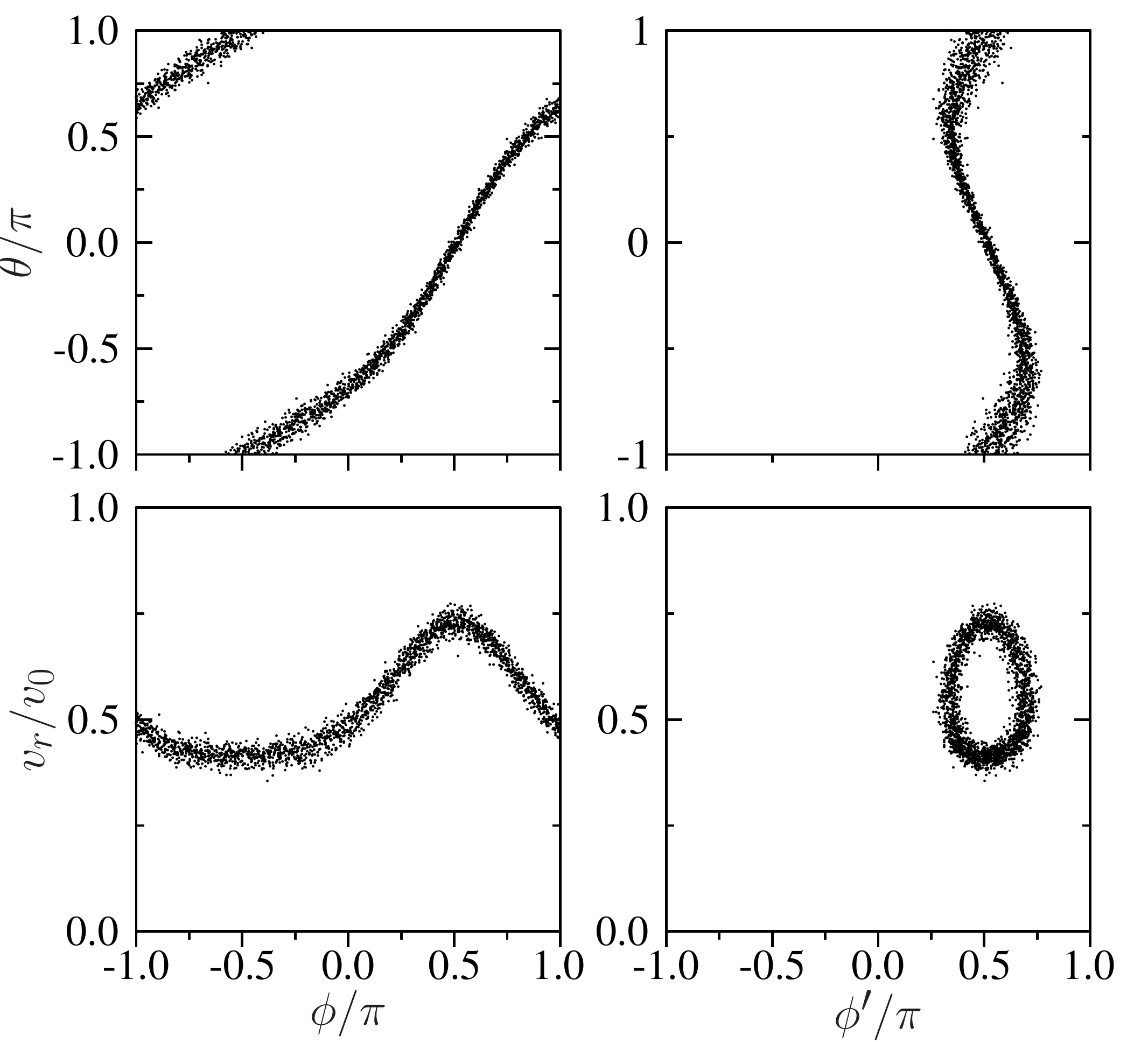}
\end{center}
\caption{ Synchronization between disk collision angle $\theta$ 
  and microwave phase $\phi$.
  \textit{Left column}:   Dependence of angle $\theta$ of 
  collision point on disk,
  counted from $x-$axis, and radial velocity $v_r$, taken at collision,  
  on microwave phase $\phi$.
  \textit{Right column}: Same as in left column with $\phi'=\phi-\theta$.
  Here $j=2.25$, $\epsilon=0.04$, $\epsilon_s=0.001$, $\gamma_d/\omega=0.01$,
  $v_0=v_F$,
  there is no noise; points are shown for times 
  $10^4/\omega < t <10^5/\omega$,  
  the capture time of this orbit is $t_c > 10^5/\omega$.
  Data for model (D1).
} 
\label{fig7}
\end{figure}

For some impact parameters a trajectory can be captured
for a very long time $t_c$, in certain cases in absence of noise we have
$t_c=\infty$. At such long capture times
the collisions with disk become synchronized with the
phase $\phi$ of microwave field at the moment of collisions.
This is directly illustrated in Fig.~\ref{fig7}
where we show the angle $\theta$ of a collision point
on disk, counted from $x-$axis, in dependence on $\phi$.
Indeed, the dependence $\theta$ on $\phi$ forms a smooth curve
corresponding to synchronization of two phases.
At the same time the radial velocity at collisions
$v_r$ moves along some smooth invariant curve $v_r(\phi)$
in the phase space $(v_r,\phi)$. 
However, to make a correct comparison with 
the radial field models (DR1), (DR2) we should take into
account that the cyclotron circle rotates around disk so that
we should draw the Poincar\'e section in the rotational
phase $\phi'=\phi-\theta$. In this representation we 
see the appearance of the resonance (see right column of Fig.~\ref{fig7})
that is similar to those seen in Figs.~\ref{fig4},~\ref{fig5}
for the radial field models.

\begin{figure}[!ht] 
\begin{center} 
\includegraphics[width=0.9\columnwidth]{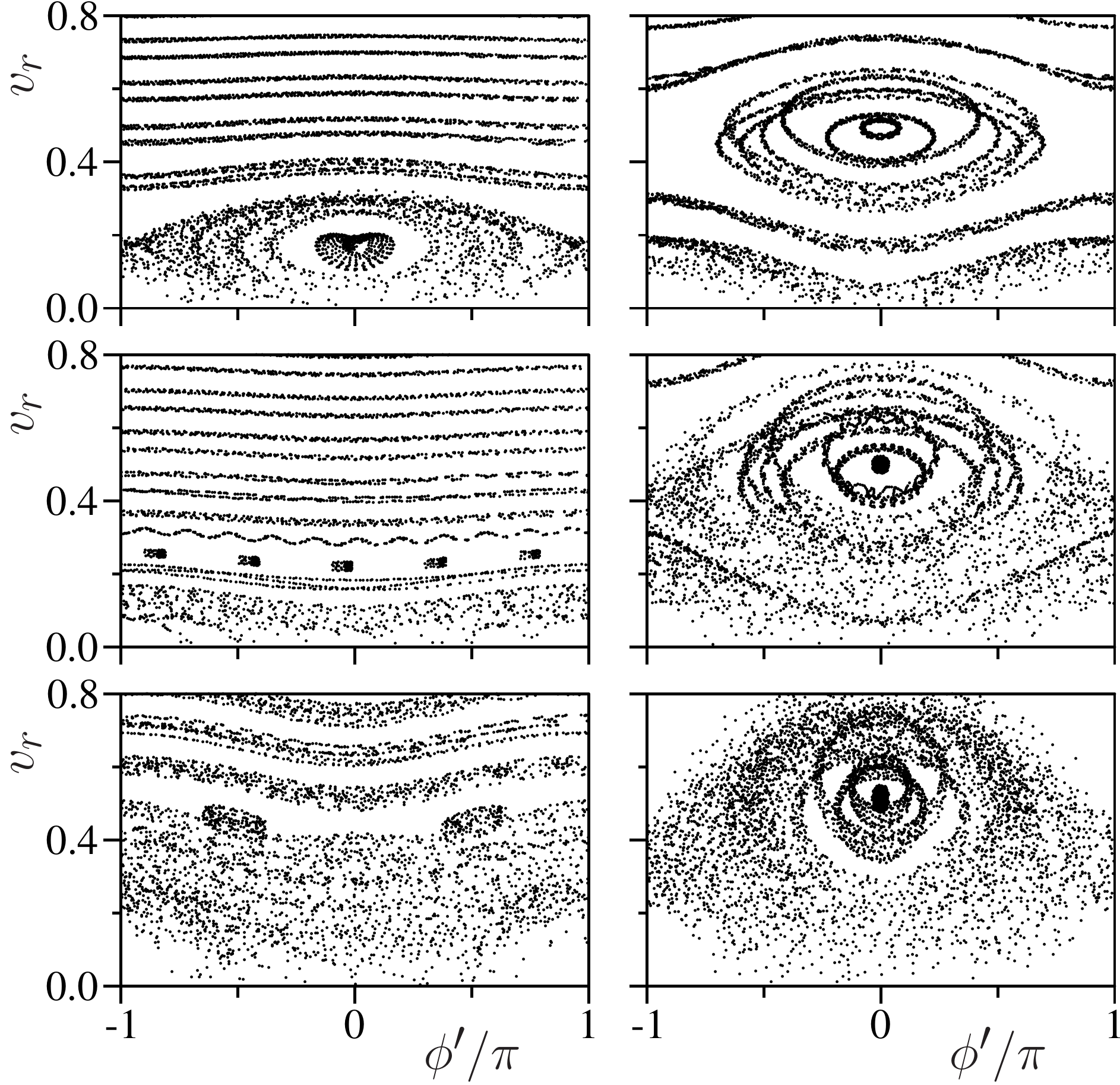}
\end{center}
\caption{ Poincar\'e sections in phase plane
  $(v_r,\phi')$ with $\phi'=\phi-\theta$
  for $j=2.1$, $\epsilon=0.01$ (left top); 
  $j=2.25$, $\epsilon=0.01$ (right top);
   $j=2.75$, $\epsilon=0.02$ (left middle);
    $j=2.25$, $\epsilon=0.02$ (right middle);
    $j=2.75$, $\epsilon=0.04$ (left bottom);
   $j=2.25$, $\epsilon=0.04$ (right bottom);
  $v_r$ is expressed in units of $v_F$.
  Data for model (D1), no noise no dissipation.
} 
\label{fig8}
\end{figure}

A more direct correspondence between radial field models
(DR1), (DR2) and the model (D1) with a linearly polarized microwave field
is well seen from the Poincar\'e sections shown in the rotation frame
of phase $\phi'=\phi-\theta$ in Fig.~\ref{fig8}. In this frame we 
see directly the resonance at $j=2.1, 2.25$ being very similar to the wall case
and the radial field model.
However, the positions of resonance at $v_r=v_{res}$ are different from those in 
Fig.~\ref{fig5}. Of course in the rotation frame
the orbital momentum is only approximately conserved 
that gives a broadening of invariant curves in Fig.~\ref{fig8}. 

We explain this as follows. For the linear polarized field
of model (D1) the radial component of microwave field
is proportional to $\epsilon_r \sim \epsilon \sin \theta \cos \omega t 
\sim 0.5 \epsilon \sin (\omega t - \theta)$ where we kept only 
slow frequency component of radial field
(the neglected term with $ \sin (\omega t + \theta)$
gives resonant values $v_{res} > v_F$). 
The radial field $\epsilon_r$ gives 
kicks to the radial velocity component
at collisions with disk similar to the case of model (DR2)
described by Eq.~(\ref{eq3}): ${\bar v_r} = v_r +0.5 \epsilon \sin \phi'$,
${\bar \phi'} = \phi' + (2\pi j - 2 {\bar v_r} j/\rho) - 
2 {\bar v_r (\rho-1)/\rho}$.
Here we use the radial field component phase
$\phi'=\omega t - \theta $ at a moment of collision with disk
(the tangent component does not change $v_r$ and can be neglected).
The phase variation ${\bar \phi'}-\phi'$
has the first term $ 2\pi j - 2 {\bar v_r} j/\rho$
being the same as for the radial field model (DR2),
and an additional term related to rotation around disk
with $-\Delta \theta =- 2 {\bar v_r} (\rho-1)/\rho$ 
which comes from geometry. Indeed, the  segment angles of intersections of 
circles $r_d$ and $r_c$ are:  for disk radius $r_d$ it is 
$\Delta \theta = 2 {\bar v_r} (\rho-1)/\rho$
and for cyclotron radius  $r_c$ it is
$\Delta \varphi = 2 {\bar v_r} /\rho$. Thus, their ratio 
is  $\Delta \theta/\Delta \varphi = r_c/r_d$ in agreement
with the geometrical scaling.
This result can be obtained from the expression for
$\Delta \varphi$ by interchange of two disks
that gives the above expression for $\Delta \theta$
(at $r_d=r_c$ both shifts   $\Delta \varphi= 2 {\bar v_r}/\rho$
and $\Delta \theta= 2 {\bar v_r} (\rho-1)/\rho$ are equal).

Thus again the dynamical description is reduced down 
to the Chirikov standard map
with slightly modified parameters giving us for the model (D1)
the chaos parameter $K=2 \epsilon (j+\rho -1)/\rho$
being usually smaller than unity, resonance position
$v_{res}$,
resonance width $\delta v$ 
and the resonance energy width $E_r=(\delta v)^2/2$:
\begin{eqnarray}
\label{eq5}
E_r & = & 8 \epsilon \rho E_F/(\rho + j -1) ; \; 
\rho=1+r_c/r_d \; ; \\
\nonumber
v_{res} &  = & \pi \rho \delta j/(j+\rho-1) ; \; 
\delta v = 4 \sqrt{\epsilon \rho/(2 (j+\rho-1))}; \\
\nonumber
\delta j_\epsilon & = & \delta v (\rho+j-1)/(2\pi \rho); 
\; j=\omega/\omega_c \; ,
\end{eqnarray}
where $\delta j_\epsilon$ is a shift of resonance produced by
a finite separatrix half width $\delta v /2$. 
For our numerical simulations we have
$\rho=1+j$ with
$v_{res}=\pi (j+1) \delta j/(2j)$,
$\delta v = 2\sqrt{\epsilon (j+1)/j}$
and $\delta j_\epsilon = (2/\pi) \sqrt{\epsilon j/(j+1)}$.

At $\rho=j+1$ Eq.~(\ref{eq5}) gives the values $v_{res}=0.232$ at $j=2.1$ 
while the numerical data of Fig.~\ref{fig8} give
$v_{res} \approx 0.2$, and we have at $j=2.25$
the theory value $v_{res}= 0.567$ being in a good agreement
with the numerical value  $v_{res} \approx 0.5$ of Fig.~\ref{fig8}. 
For $j=2.75$ we have the resonance position
at $v_r<0$ corresponding to the bulk and thus the resonance is absent. 
The resonance width
in Fig.~\ref{fig8} at $j=2.25$, $\epsilon=0.01$ can be
estimated as $\delta v \approx 0.3$ that is in a satisfactory
agreement with the theoretical value $\delta v =0.24$ from (\ref{eq5}).
We remind that in model (D1) we have only approximate
conservation of orbital momentum
that gives a broadening of invariant curves
and makes determination of the resonance width less accurate.
In spite of this broadening we see that the resonance description
by the Chirikov standard map works rather well.

\begin{figure}[!ht] 
\begin{center} 
\includegraphics[width=0.9\columnwidth]{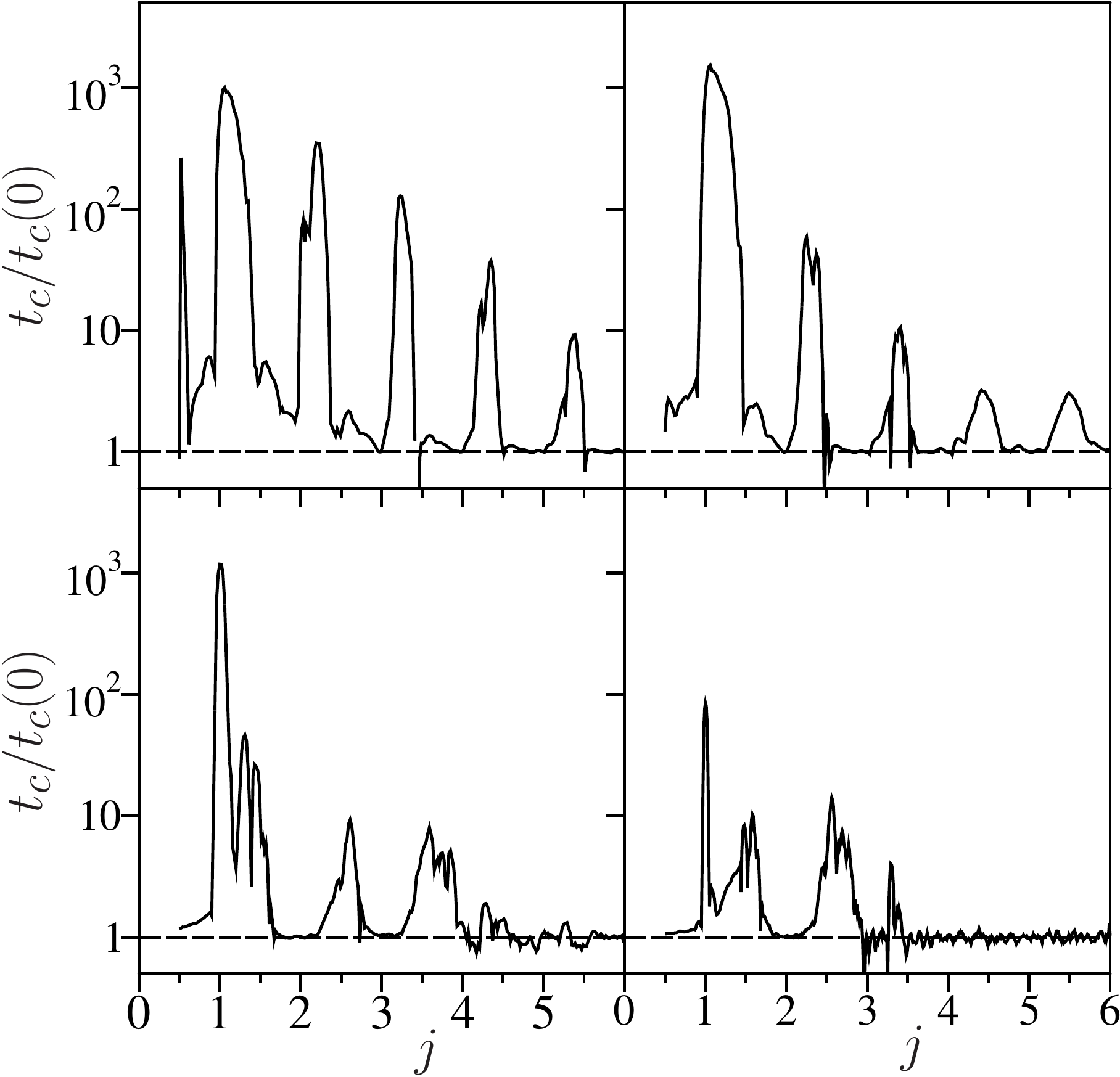}
\end{center}
\caption{ Dependence of rescaled
  capture time $t_c/t_c(0)$ on $j$ at $\epsilon=0.04$ 
  shown at various static fields: 
  $\epsilon_s=2.5\cdot 10^{-4}$ (top left),
  $ 10^{-3}$ (top right), $ 8\cdot 10^{-3}$ (bottom left),
  $0.016$ (bottom right). Here $\gamma_d/\omega=0.01$, there is no noise.
  Data for model (D1).
} 
\label{fig9}
\end{figure}

In Fig.~\ref{fig9} we show the dependence of average
capture time $t_c$ on $j$ in the model (D1). The averaging is done 
over $N_s=500$ trajectories scattered on disk 
at all such impact parameters that cyclotron orbit 
can touch the disk. Here, we show the ratio $t_c/t_c(0)$
where $t_c(0)$ is an average capture time in absence of microwave.
According to our numerical data we have approximate
dependence $\omega t_c(0) \approx 3/\sqrt{\epsilon_s}$
corresponding to a period of nonlinear oscillations
in a disk map description discussed in \cite{berglund,aliktrap}.

The data of Fig.~\ref{fig9} show a clear periodic dependence
of capture time $t_c$ on $j$ corresponding to the periodicity
variation of Poincar\'e section with $j$ 
(see Figs.~\ref{fig4},~\ref{fig5}.~\ref{fig8}). 
This structure is especially well visible
at weak static fields. 
With an increase of $\epsilon_s$
this structure is suppressed. Indeed, at large $\epsilon_s$
even without microwave field the trajectories can escape
from disk as it discussed in \cite{berglund,aliktrap}
and microwave field does not affect the scattering in this regime.

\begin{figure}[!ht] 
\begin{center} 
\includegraphics[width=0.9\columnwidth]{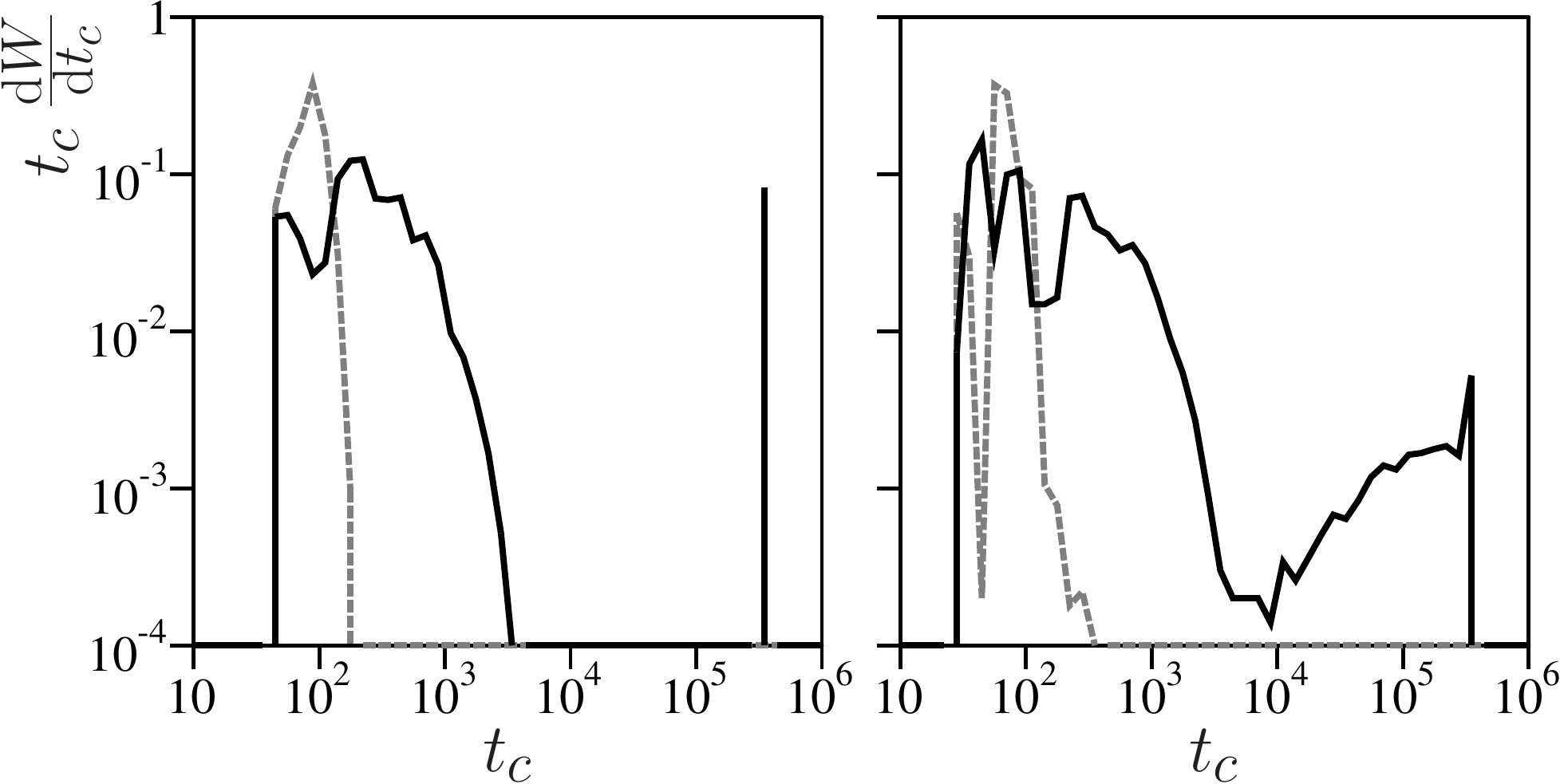}
\end{center}
\caption{ 
  Differential probability distribution $t_cdW/dt_c$
   of capture times $t_c$ for parameters of Fig.~\ref{fig9}.
   {\it Left panel:} 
   $j=2$ (dashed curve for minimum of capturing probability) 
   and $j=2.25$ (full curve at maximum of  capturing probability)
   at $\epsilon_s=0.001$. {\it Right panel:}
   similar cases at  $j=1.9$ (dashed curve) and
   $j=2.37$ (full curve) at   $\epsilon_s=0.002$.
   Data are obtained with $N_s= 5\cdot 10^4$ 
   trajectories started at different impact parameters 
   and running up to time $t=3\cdot 10^5/\omega$.
   Here $t_c$ is expressed in units of $1/\omega$.
   Data for model (D1).
} 
\label{fig10}
\end{figure}

The distributions of capture times are shown in Fig.~\ref{fig10}.
We clearly see that at resonant values
of $j$ a microwave field leads to appearance of long capture times.
For example,  we have the probability to be captured for 
$t_c > 180/\omega$ being $W=0.46$ at $j=2.25$ 
while at $j=2$ we have $W=0$(left panel in Fig.~\ref{fig10}); 
and we have   $W=0.38$ at $j=2.37$ 
 while at $j=1.9$ we have $W < 3 \cdot 10^{-4}$ 
(right panel in Fig.~\ref{fig10}).
These data confirm much stronger capture at certain resonant
values of $j$.

\begin{figure}[!ht] 
\begin{center} 
\includegraphics[width=0.9\columnwidth]{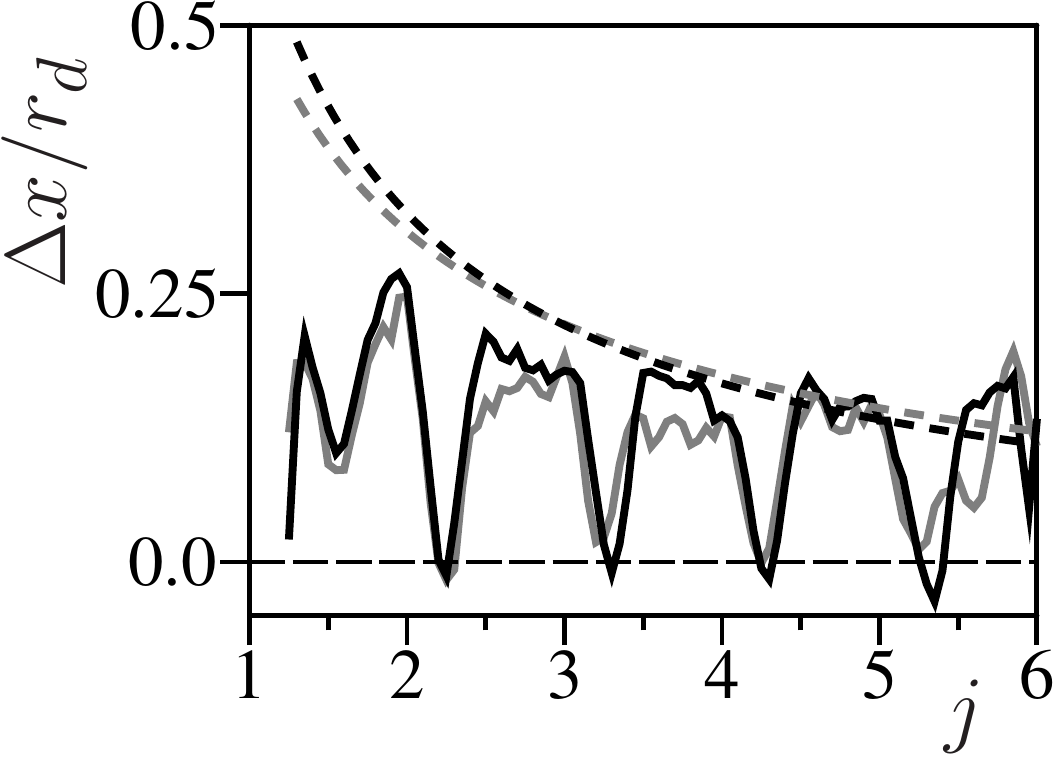}
\end{center}
\caption{ 
  Average shift $\Delta x$ along static 
  field after scattering on a single disk shown as a function of $j$
  at various amplitudes of static and microwave fields:
  $\epsilon_s=0.0005$, $\epsilon=0$ (gray dashed curve);
  $\epsilon_s=0.001$, $\epsilon=0$ (black dashed curve);
  $\epsilon_s=0.0005$, $\epsilon=0.04$ (gray full curve);
  $\epsilon_s=0.001$, $\epsilon=0.04$ (black full curve).
  The data are obtained by averaging over $N_s=5\cdot 10^3$
  scattered trajectories with random impact parameters;
  here $\gamma_d/\omega=0.01$, noise amplitude $\alpha_i=0.005$.
  Data for model (D1).
} 
\label{fig11}
\end{figure}

The data of Figs.~\ref{fig9},~\ref{fig10} show that 
the scattering process on disk is strongly modified by a microwave field.
However, to determine the conductivity properties of a sample
we need to know what is an average displacement $\Delta x$
along static field after a scattering on a single disk. 
Indeed, in our model
a dissipation is present only during collisions with disk
while in a free space between disks the dynamics is integrable and
Hamiltonian. Hence during such a free space motion
there is no displacement along the static field
(the dissipative part of conductivity or resistivity
appears only due to dissipation on disk).
The dependence of $\Delta x$ on $j$ is shown in Fig.~\ref{fig11}.
In absence of microwave field at $\epsilon =0$ we find
$\Delta x \propto 1/j \propto B$ that corresponds to a simple estimate
$\Delta x \propto \omega_c \gamma_d$. The numerical data
show that $\Delta x$ is practically independent of 
$\epsilon_s$ and that $\Delta x =0$ in absence of dissipation at
$\gamma_d=0$. In presence of microwave field we see that
the displacement along the static field
has strong periodic oscillations with $j$.
The striking feature of Fig.~\ref{fig11} is the
appearance of windows of zero displacement $\Delta x \approx 0$ at
resonance values $j_r=9/4, 13/4,  17/4 ...$.
We discuss how this scattering on a single disk modifies
the resistance of a sample with large number of disks
in next Section.

\section{Resistance of samples with many disks}

To determine a resistance of a sample with many disks
we use the following scattering disk model.
The scattering on a single disk 
in a static electric field $\epsilon_s$
is computed as it is described
in the previous Section with a random impact parameter
inside the collision  cross section
$\sigma_d=2(r_c+r_d)$. After that a trajectory
evolves along $y$-axis 
according to the exact solution of Hamiltonian Eq.~(\ref{eq1}) 
(no dissipation and no noise) up to a 
collision with next disk which is taken randomly
on a distance between $2(r_c+r_d)$ and $2 \ell_e$
where $\ell_e=1/(\sigma_d n_d)$ is a mean free path
along $y-$axis and $n_d$ is a two-dimensional density
of disks (of course $\ell_e \gg 2(r_c+r_d)$). 
In a vicinity of disk the dynamical evolution
is obtained by Runge-Kutta solution of dynamical equations
as it was the case in previous Section.
We use low disk density with $n_d r_d^2 \sim 1/100$.
The collision with disk is done with a random impact parameter
in the $x-$axis of disk vicinity: the impact parameter
is taken randomly in the interval $[-(r_c+r_d),(r_c+r_d)]$
around disk center. 
Noise acts only when a center of cyclotron radius of trajectory is 
on a distance $r<r_d+r_c$ from disk center 
so that a collision with disk is possible.
After scattering on a disk a free propagation follows 
up to next collision with disk. 

Along such a trajectory we
compute the average displacement $\delta x$ and $\delta y$ after 
a time interval $\delta t$.
In this way the number of collisions with disks is
$N_{col} \approx \delta y/\ell_e$ and a total displacement in $x-$axis is
$\delta x \approx N_c \Delta x$ where $\Delta x$ is an average
displacement on one disk discussed in previous Section.
We compute the global displacements $\delta x, \delta y$
on a time interval $\delta t =10^6/\omega$  averaging data
over 200 trajectories. We call this system model (D2).
\begin{figure}[!ht] 
\begin{center} 
\includegraphics[width=0.9\columnwidth]{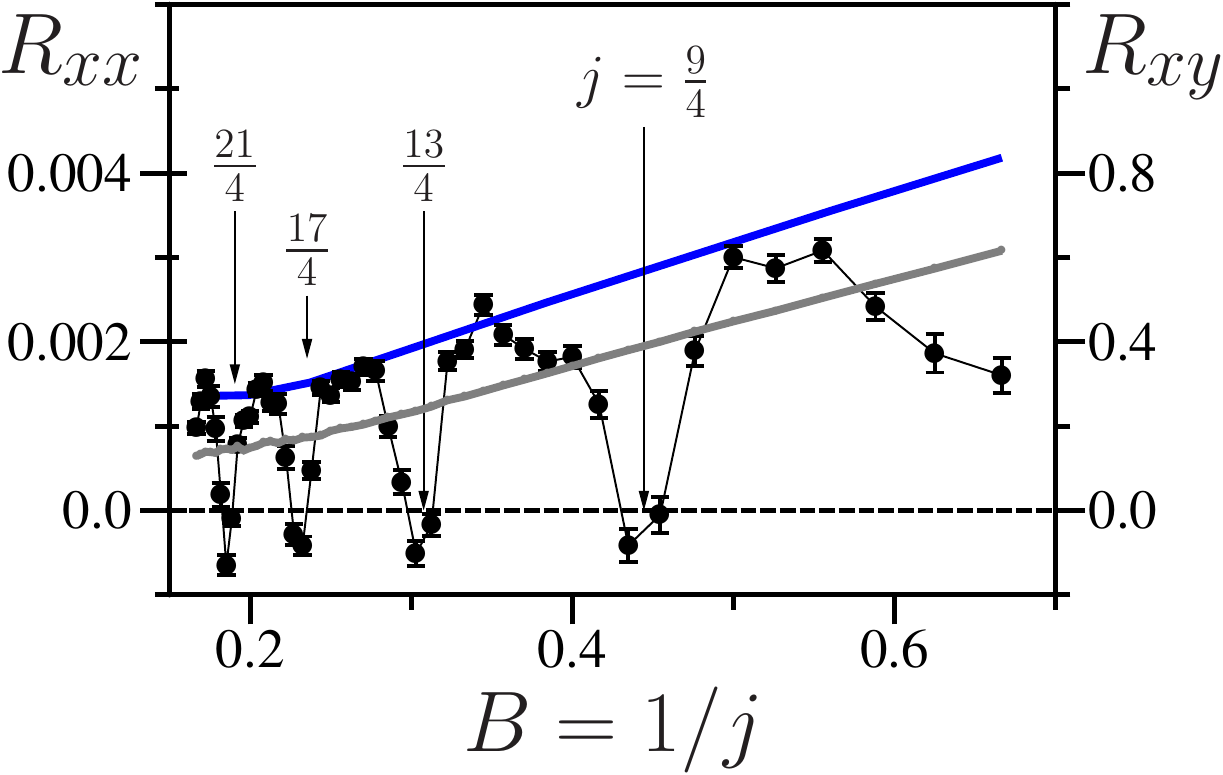}
\end{center}
\caption{Dependence of resistivity  
  $R_{xx}$ and $R_{xy}$ on magnetic field $B=1/j$
  (resistivity is expressed in arbitrary numerical units).
  Blue and gray curves show respectively $R_{xx}$ and $R_{xy}$
  in absence of microwave field. Black curve with points show 
  $R_{xx}$ dependence of $B$ at microwave field $\epsilon=0.04$
  (here $B$ is expressed in units of $1/\omega=const$).
  Here $\epsilon_s=0.001$, $\gamma_d/\omega=0.01$, noise amplitude 
  $\alpha_i=0.005$, $\tau_i=1/\omega$. 
  Resonant values $j_r$ are shown by arrows;
  bars show statistical errors for $R_{xx}$. Data are obtained by averaging 
  over 200 trajectories propagating up to time $t=10^6/\omega$.
  Data for model (D2).
} 
\label{fig12}
\end{figure}

Then the current components are equal to
$j_x= \delta x/\delta t$, $j_y= \delta y/\delta t$
and conductivity components are $\sigma_{xx}=j_x/E_{dc}$,
$\sigma_{xy}=j_y/E_{dc}$ (the current is computed per one electron).
We work in the regime of weak {\it dc-}field
where $j_x, j_y$ scales linearly with $E_{dc}$.
The current $j_y$ is determined by the drift velocity
$v_d=E_{dc}/B \ll v_F$.
Since the mean free path is large
compared to disk size  
$\ell_e \gg r_c \ge r_d$ we have 
an approximate relation $\sigma_{xy} \approx 1/B$, 
$\sigma_{xx} \approx  \Delta x /B\ell_e \approx \sigma_{xy} \Delta x/\ell_e$.
As in 2DEG experiments \cite{mani2002,zudov2003} we have in our simulations
$\sigma_{xy}/\sigma_{xx} = R_{xy}/R_{xx} \sim 100$ (see Fig.~\ref{fig12}).
The resistivity is obtained by the usual inversion of conductivity
tensor with $R_{xx}=\sigma_{xx}/(\sigma^2_{xx}+\sigma^2_{xy}) \approx \sigma_{xx}/\sigma^2_{xy}$,
$R_{xy}=\sigma_{xy}/(\sigma^2_{xx}+\sigma^2_{xy}) \approx 1/\sigma_{xy}$.
The dependence of $R_{xx}$, $R_{xy}$, expressed in arbitrary numerical units,
on magnetic field $B=\omega/j=1/j$ is shown in Fig.~\ref{fig12}. 

In absence of microwave field
we find $R_{xy} \propto B$ and $R_{xy}/R_{xx} \approx 200$
similar to experiments \cite{mani2002,zudov2003}.
For small noise amplitude (e.g. $\alpha_i=0.005$) we have $R_{xx}$ growing
linearly with $B$ (see Fig.~\ref{fig12}) 
but at larger amplitudes (e.g. $\alpha_i =0.02$)
its increase with $B$  becomes practically flat showing only $30 \%$ increase
in a give range of $B$ variation.

\begin{figure}[!ht] 
\begin{center} 
\includegraphics[width=0.9\columnwidth]{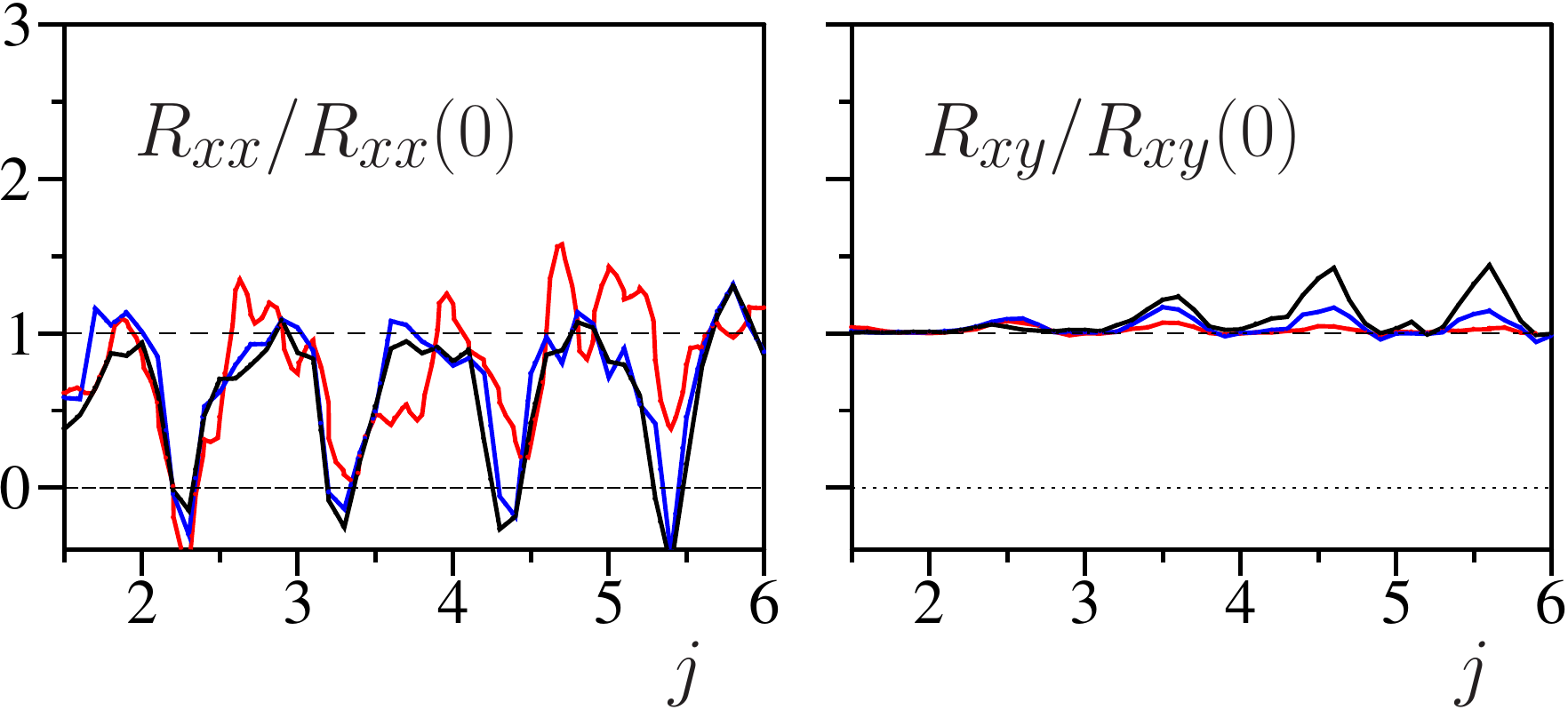}
\end{center}
\caption{(Color online) Rescaled values of resistivity
  $R_{xx}$ (left panel) and 
  $R_{xy}$ (right panel) as function of $j=\omega/\omega_c$
  at various noise amplitudes
  $\alpha_i=0.005$ (black curve),
  $0.01$ (blue/dark curve), $0.02$ (red/gray curve).
  Here $\epsilon=0.04$, $\epsilon_s=0.001$,
   other parameters are as in Fig.~\ref{fig12}.
   Curves are drown though numerical points
   obtained with a step $\Delta j =0.1$. Data for model (D2).
} 
\label{fig13}
\end{figure}

\begin{figure}[!ht] 
\begin{center} 
\includegraphics[width=0.9\columnwidth]{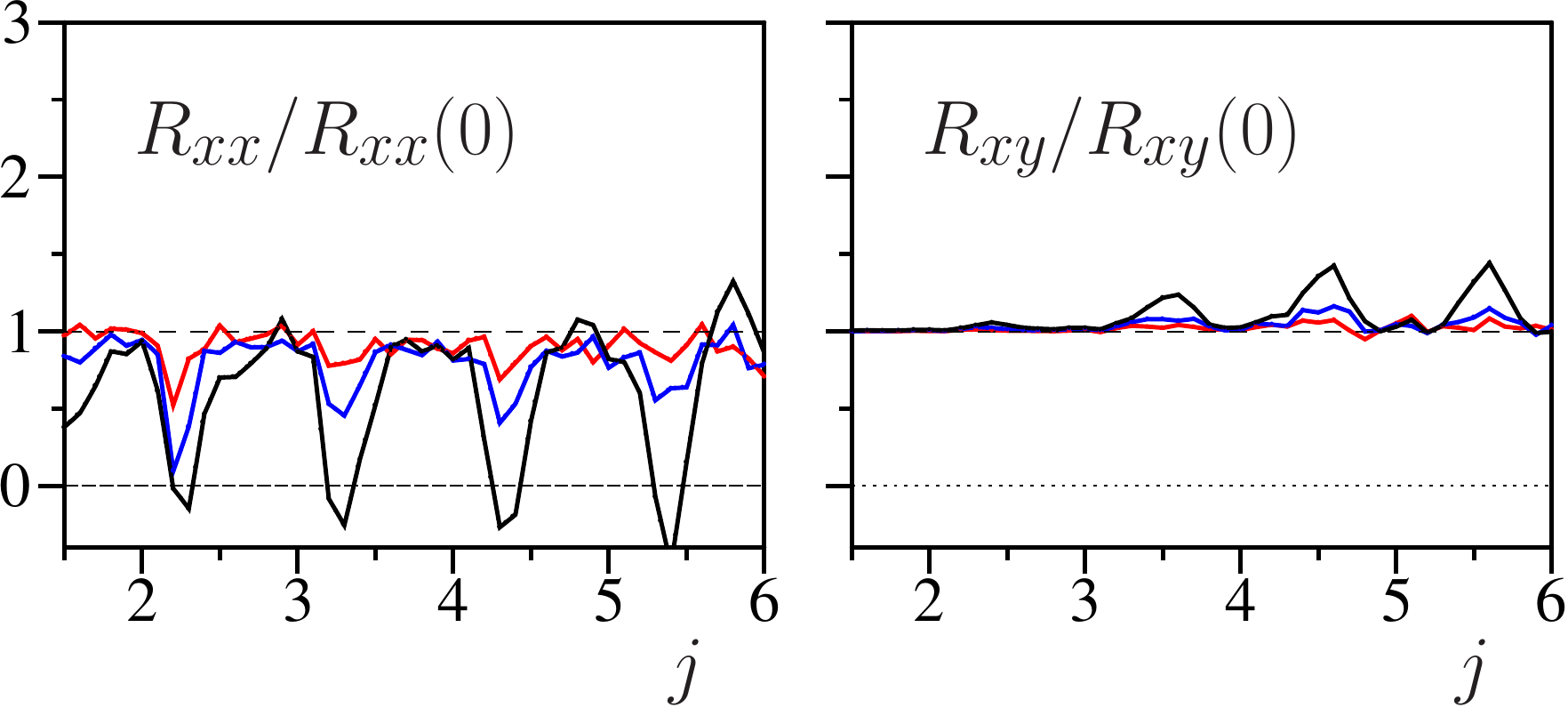}
\end{center}
\caption{(Color online) Same as in Fig.~\ref{fig13} at various
  microwave amplitudes $\epsilon=0.01$ (red/gray curve),
  $0.02$ (blue/dark curve), $0.04$ (black curve);
  amplitude of noise is fixed at $\alpha_i=0.005$. Data for model (D2).
} 
\label{fig14}
\end{figure}

In presence of microwave field the dependence of $R_{xx}$ on $B$
is characterized by periodic oscillations with minimal 
$R_{xx}$ values being close to zero at resonant values of $j=j_r$
well visible in Fig.~\ref{fig12}. The dependence of $R_{xx}$, $R_{xy}$
rescaled to their values $R_{xx}(0)$, $R_{xy}(0)$ in absence of 
microwave field are shown in Fig.~\ref{fig13} at various amplitudes
of noise and fixed $\epsilon$, and in Fig.~\ref{fig14} 
at various $\epsilon$ and fixed noise amplitude $\alpha_i$. 
We see that increase on noise leads to an increase of minimal
values of $R_{xx}$ at resonant values $j_r$.
In a similar way a decrease of microwave power leads to increase
of minimal values of  $R_{xx}$ at $j_r$. At the same time
the Hall resistance $R_{xy}$ is only weakly affected by 
microwave radiation as it also happens in ZRS experiments.

These results are in a qualitative agreement with the 
ZRS experiments. On the basis of our numerical studies
we attribute the appearance of approximately zero resistance
at $j_r$ values in our bulk model of disk scatterers
to  long capture times of orbits 
in disk vicinity at these $j_r$ values (see Fig.~\ref{fig9}).
During this time $t_c$ noise gives fluctuations
of collisional phase $\theta$ and due to that
a cyclotron circle escapes from disk practically 
at random displacement $\Delta x$ that after averaging 
gives average $\Delta x =0$. Since resistivity
is determined by the average value of $\Delta x$
this leads to appearance of ZRS.
We note that this mechanism is 
different from the one of edge transport stabilization discussed 
here and in \cite{adcdls}. However, both mechanisms
are related to a long capture times near  edge or near disk
that happens due to synchronization of cyclotron phase with
microwave field phase and capture inside the nonlinear resonance.
\begin{figure}[!ht] 
\begin{center} 
\includegraphics[width=0.9\columnwidth]{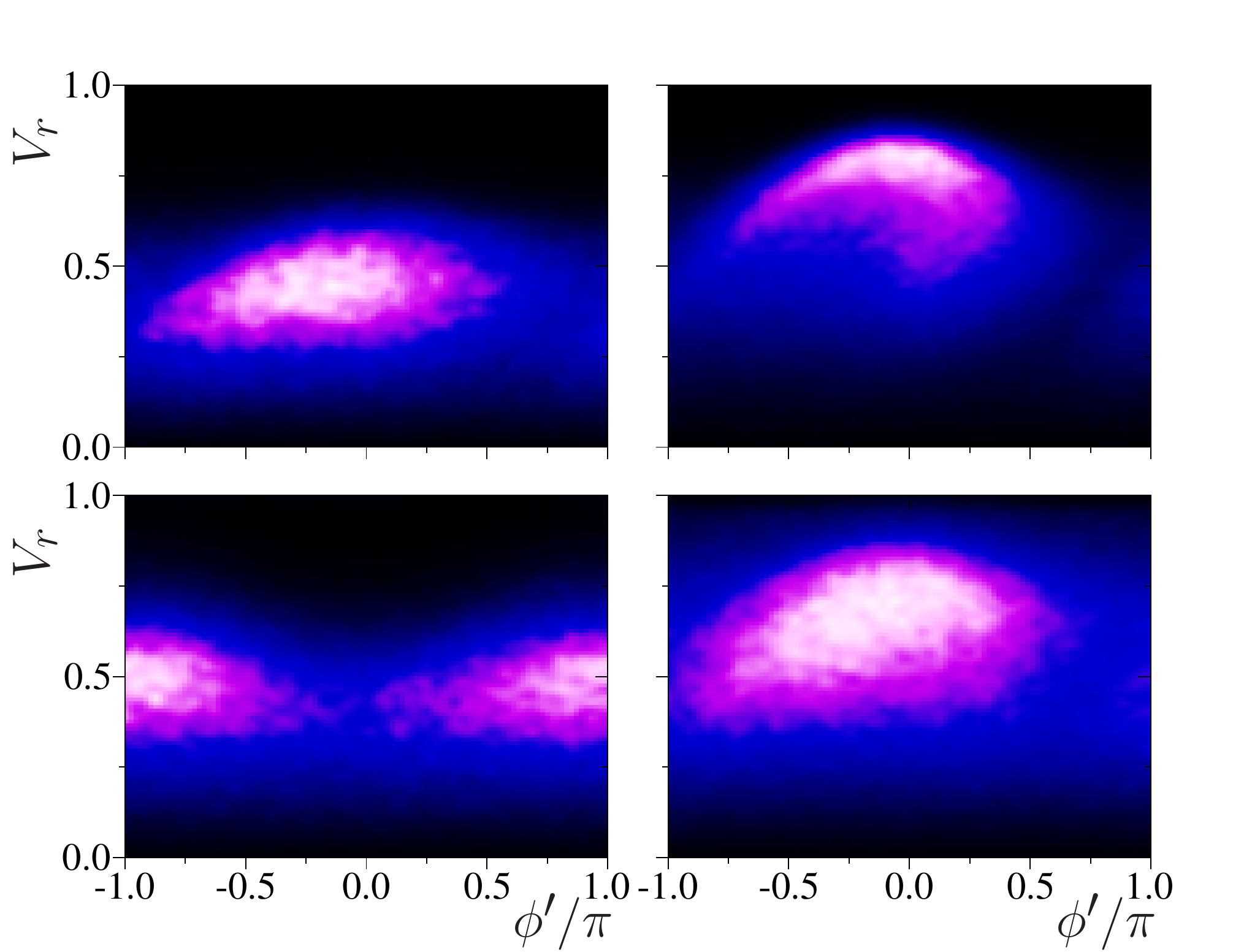}
\end{center}
\caption{(Color online) 
   Phase space $(v_r,\phi')$ of trajectories at the moment of collisions 
   with disks for parameters of Fig.~\ref{fig14} at $\epsilon=0.04$:
   $j=2.1$ (top left), $j=2.25$ (top right),
   $j=2.75$ (bottom left), $j=3.25$ (bottom right).
   Here $\phi'=\phi-\theta$ where $\phi=\omega t$ is a microwave phase
   at the moments of collisions with disk and $\theta$ is the angle 
   on disk at collision moment, counted from $x$-axis 
   (same as in  Figs.~\ref{fig7},~\ref{fig8}).
   Data are obtained from $500$ trajectories
   iterated up to time $t=10^6/\omega$. 
   Density of points
   is shown by color with black at zero and white at maximum density.
   The average number of collisions per disk per trajectory is
   $N_{col}=12.4$, $25.9$, $9.5$, $15.5$ respectively for
   $j=2.1$, $2.25$, $2.75$, $3.75$.
   Data are obtained for model (D2).
     }
\label{fig15}
\end{figure}

To illustrate the capture inside the resonance we 
present the distributions of 
trajectories from Fig.~\ref{fig14}
shown in the phase space plane $(v_r,\phi')$
at the moments of collisions with disks in Fig.~\ref{fig15}.
This is similar to the Poincar\'e sections of Fig.~\ref{fig8}
however, now we consider the real case of diffusion and scattering on many
disks in the model (D2) with noise and dissipation. 
We see that for $j=2.25$ orbits are captured
in a vicinity of the center of nonlinear resonance
at $\phi' \approx 0$ 
well seen in Fig.~\ref{fig8}.
For $j=2.1$ we have a density maximum located at smaller values of $v_{res}$
and $\phi' \approx 0$
even if there is a certain shift of $v_{res}$ 
produced by a significant resonance width
at $\epsilon=0.04$. At $j=2.75$ we have a density maximum at
$\phi' \approx \pm \pi$  corresponding to an unstable fixed point of
separatrix. The total number of collision points
$N_{col}$ in this case is by a factor $2.5$ smaller than in the case 
of stable fixed point at $j=2.25$. A similar 
situation is seen in the case of wall model (W1)
(see Fig.1d,f in \cite{adcdls}) even if there
the ratio between number of captured points
was significantly larger.
The results of Fig.~\ref{fig15}
show that in the ZRS phase the collisions with disk
indeed create synchronization of cyclotron and microwave phases and capture
of trajectories inside the nonlinear resonance.

However, there are also some distinctions between bulk disk model (D2)
and experimental observations.
The first one is that there are minima for $R_{xx}/R_{xx}(0)$
but there are no peaks which are well visible around integer
$j$ values in ZRS experiments \cite{mani2002,zudov2003},\cite{dmitrievrmp}
and numerical simulations of transport along the edge \cite{adcdls}.
The second one is appearance of small negative 
values of $R_{xx}$ at $j_r$ values.

\begin{figure}[!ht] 
\begin{center} 
\includegraphics[width=0.9\columnwidth]{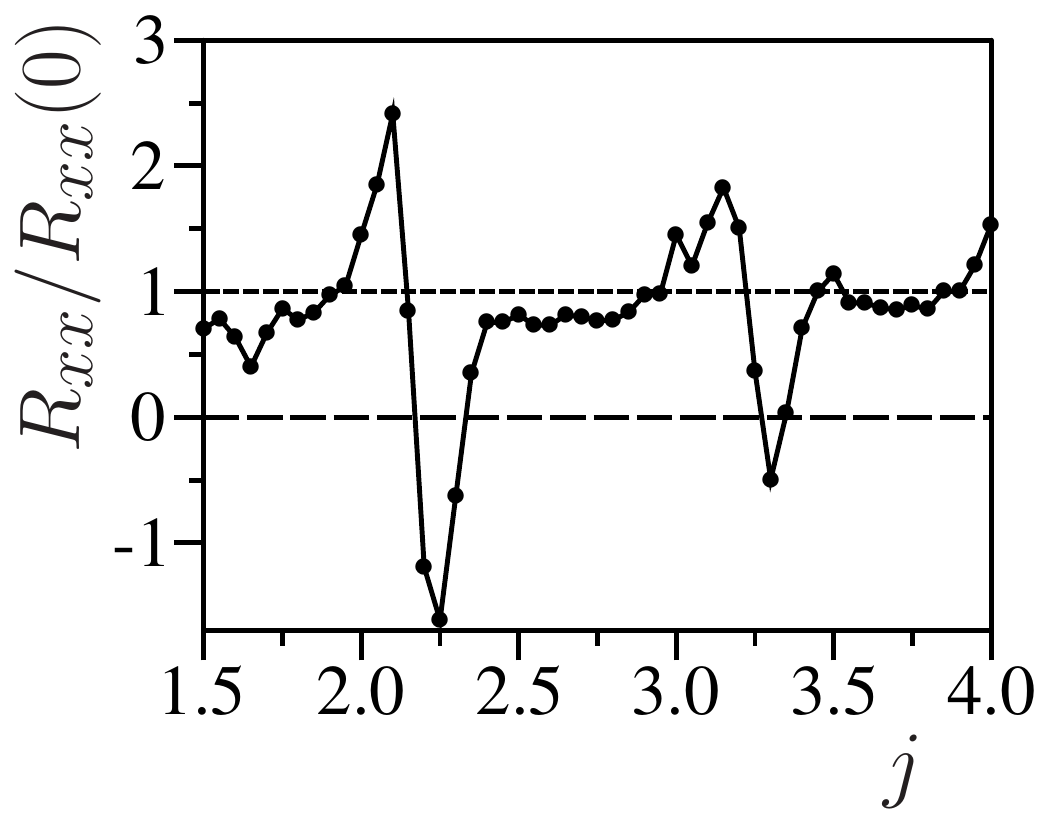}
\end{center}
\caption{Dependence of $R_{xx}/R_{xx}(0)$ on $j$
  in the disk model with dissipation at disk collisions
  at rate $\gamma_d/\omega =0.01 $ and dissipation in disk vicinity
  with rate $\gamma_0/\omega=0.02$;
  a disk roughness  gives additional angle rotations
  with amplitude $\alpha_d=0.1$ (see text);
  the amplitude of noise in disk vicinity is $\alpha_i=0.001$.
  Here we have $\epsilon=0.04$, $\epsilon_s=0.001$; 51 numerical points
  in $j$ are connected by lines to adapt an eye.  
  Data are obtained by averaging 
  over 100 trajectories propagating up to time $t=10^6/\omega$.
  Data for model (D3).}
\label{fig16}
\end{figure}

We attribute the absence of peaks to a specific dissipation mechanism
which takes place only at disk collisions. It is rather convenient 
to run long trajectories using exact solution for free
propagation between disks. Indeed, in this scheme 
there is no dissipation during this free space
propagation and thus these parts of trajectories have no
displacement along static field. We also tested 
a dissipation model with additional $\gamma(v)= \gamma_0 (|v/v_F|^2-1)$
for $|v| > v_F$ and $\gamma(v)=0 $ for $|v| \leq v_F$.
This dissipation works only in a disk vicinity
when the distance between disk center and cyclotron center 
is smaller than $r_d+r_c$. The dissipation $\gamma_d$ 
on disk remains unchanged.
We also added a certain roughness of disk surface
modeled as an additional random angle rotation 
of velocity vector in the range $\pm \alpha_d$, done
at the moment of collision with disk. 
We call this system disk model (D3).
The results
for the resistivity ratio  $R_{xx}/R_{xx}(0)$ 
are shown in Fig.~\ref{fig16}.
They show an appearance of clear peaks of $R_{xx}/R_{xx}(0)$ 
in presence of
such additional dissipation in vicinity of integer $j$.
There is also a small shift of minima from integer plus $1/4$
to integer plus $0.4 - 0.5$.

The second point of distinctions from ZRS experiments
is a small negative value of $R_{xx}$ at resonant $j$
values. It is relatively small  for disk model (D2) 
(see Figs.~\ref{fig13},~\ref{fig14}) and 
it becomes more pronounced for disk model (D3)
(see Fig.~\ref{fig16}). 
It is possible that a scattering on disk
in presence of dissipation, noise, static and microwave
field gives a negative displacement 
$\Delta x$ which generates such negative $R_{xx}$ values.
We expect that in the limit of static field going to zero
this effect disappears. Indeed, the negative values
become smaller at smaller $\epsilon_s$ according to
data of Fig.~\ref{fig11} 
but unfortunately the small $\epsilon_s$ limit 
is also very difficult to investigate numerically.

We consider that at this stage of the theory 
the presence of negatives values for $R_{xx}$ 
does not constitute a critical disagreement. 
The escape parameters for electrons that have been captured 
on an impurity for a time long enough to make many rotations around it, 
are likely to strongly depend on the model 
for the electron impurity interaction
and further theoretical work on a more microscopic model is needed.
In general a zero average displacement along 
the field direction seems natural
for a smooth distribution of trapping times 
with a characteristic time scale much larger 
than the rotation time around 
the impurity (this assumption does not seem to hold for our 
model, see for example the sharp features on Fig.~10).
Finally in Section VII we propose a slightly 
different mechanism by which the combination 
of trapping on impurities investigated 
here and electron-electron interactions 
can lead to ZRS.

\section{Physical scales of ZRS effect}

The ZRS experiments \cite{mani2002,zudov2003} show that
the resistance $R_{xx}$ in the ZRS minima scales 
according to Arrehenius law 
$R_{xx} \propto \exp(-T_0/T)$ with a certain energy scale
dependent on a strength of microwave field.
In typical experimental conditions
one finds very large $T_0 \approx 20 K$ at 
$j_r=5/4$ (see e.g. Fig.3 in \cite{zudov2003}).
These data also indicate the dependence $T_0 \propto 1/j_r \propto B$.
This energy scale $k_BT_0$ is very large being
only by a factor $7$ smaller than the Fermi energy
$E_F/k_B \approx 150 K$.
At the same time the amplitude of microwave field
is rather weak corresponding to
$\epsilon \approx 0.003$ at field of $1 V/cm$
or ten times larger at $10 V/cm$
(unfortunately it is not known what is
an amplitude of microwave field acting on an electron).

As in \cite{adcdls} we argue that the Arrehenius scale 
is determined by the energy resonance width (\ref{eq4})
with $T_0=E_r/k_B$. Indeed, the resonance forms an energy barrier
for a particle trapped inside the resonance by dissipative effects
being analogous to a wash-board potential.
An escape from this potential well
requires to overcome the energy $E_r$ leading to
the Arrehenius law for $R_{xx}$ dependence on temperature.
Assuming the case of the wall with $\rho=1$
we obtain at $E=3 V/cm$ the activation temperature 
$T_0 \approx 23 K$ being is a satisfactory agreement with the
experimental observations. The theoretical relation (\ref{eq4})
also reproduces the experimental dependence $T_0 \propto 1/j_r$
at $\rho=1$. In his relation $T_0 \propto \epsilon \propto E$
being confirmed by the numerical simulations
presented in \cite{adcdls}.
This dependence is in a satisfactory agreement with
the power dependence found in experiments \cite{mani2002}.
In other samples one finds that the dependence $T_0 \propto \epsilon^2$
works in a better way. We think that higher terms
in a nonlinear resonance can be responsible for 
scaling $T_0 \propto \epsilon^2$ being different from the relation
(\ref{eq4}). Also a finite rigidity of the wall or disk scatterers
can be responsible for appearance of higher power of $\epsilon$.

The energy scale $E_r$ on disks is enhanced by a factor
$\rho=1+r_c/r_d$ for the case of radial field (\ref{eq4}). 
However, we showed that for a linear polarization 
the scale $E_r$ is given by Eq.(\ref{eq5})
and thus there is no enhancement at large $\rho$.
Indeed, we performed direct simulations at parameters
of Fig.~\ref{fig14} with the reduced value of disk radius
by a factor 2. The numerical data give approximately
the same traces $R_{xx}/R_{xx}(0)$ vs. $j$
at $\epsilon=0.01, 0.02, 0.04$ without visible
signs of deeper minima at small $\epsilon$.
This confirms the theoretical expressions (\ref{eq5}).
In any case, for small values $r_d \ll r_c$
one should analyze the quantum scattering problem which 
is significantly  more complex compared to the classical case. 
We may assume that in a quantum case one should replace 
$r_d$ by a magnetic length 
$a_B \approx r_c/\sqrt{\nu} \propto \sqrt{B}$.
In such a case we are getting $\rho =1+\sqrt{\nu} \approx 9$
that gives $T_0 \approx 8 K$ at $j \approx 2.25$ and  
microwave amplitude $E \approx 3V/cm$.
However, in this case we obtain the scale
$T_0$ being practically independent of $j$ 
which differs from experimental data.
In any case in experiments the size of impurities is
small compared to $r_c$ and a quantum treatment is 
required to reproduce the correct picture for $R_{xx}$
dependence on parameters in the ZRS phase.

Another point is related to the positions of ZRS minima
on $j$ axis. We remind that that for the wall model
the resonance is located at 
$v_{res}=\pi  \delta j/j$ (\ref{eq4}) and
that the separatrix width is $\delta v =4\sqrt{\epsilon /j}$.
The capture of trajectories from the bulk is most
efficient when a half width of separatrix
touch the border of bulk at $v_r =0$ with $v_{res}=\delta v/2$
that gives the expression $\delta j_\epsilon = 2 \sqrt{\epsilon j}/\pi$
for the wall case.
At $\epsilon =0.06$, $j=2.25$ this gives $\delta j_\epsilon =0.22$ being
in a good agreement with the numerical data $\delta j \approx 1/4$
for $R_{xx}$ dependence
on $j$ (see Figs.2,3 in \cite{adcdls} with a visible tendency of 
$\delta j$ growth with $j$). For the data presented here in Fig.~\ref{fig14}
for the disk case at $\epsilon=0.04$, $\rho=1+j$, $j=2.25$ we obtain 
from (\ref{eq5}) $\delta j_\epsilon =0.11$
that is slightly less than the numerical value $\delta j \approx 0.25$
for minima location.
We attribute this difference to an approximate nature
of expression for the resonance width at 
relatively strong microwave fields.
We also note that in experiments
an additional contribution to the value of $\delta j$ can appear
due to a finite rigidity of  disk and wall potentials.

\section{Theoretical predictions for ZRS experiments}

The theoretical models presented here and in \cite{adcdls}
reproduce main experimental features of ZRS experiments
\cite{mani2002,zudov2003},\cite{dmitrievrmp}. However, it would be useful
to have some additional theoretical predictions
which can be tested experimentally. A certain characteristic feature 
of both wall and disk models is appearance of nonlinear resonance.
For example, according to the wall model (W2) described by Eq.~(\ref{eq3})
the dynamics inside the resonance is very similar to dynamics of a pendulum.
The frequency of phase oscillations inside the resonance is
$\Omega_{ph} =\sqrt{K}= 2 \sqrt{\epsilon \omega/\omega_c} 
= 2 \sqrt{\epsilon j} \ll 1$ 
\cite{chirikov}. Here the frequency is expressed in 
number of map iterations and since the time between collisions
is approximately $2\pi/\omega_c$ we obtain the
physical frequency $\Omega_r$ of these resonant oscillations being
$\Omega_r/\omega = \Omega_{ph} \omega_c/(2\pi \omega) = 
\sqrt{\epsilon \omega_c/\omega}/\pi$.
At $\epsilon \sim 0.02$ this frequency 
is significantly smaller than the driving microwave frequency.
The dynamics inside the resonance should be
very sensitive to perturbations at frequency $\omega_1 \approx \Omega_r$
that gives:
\begin{equation}
\omega_1/\omega=\sqrt{\epsilon \omega_c/\omega} \; /\pi \;\; .
\label{eq6}
\end{equation}

To check this theoretical expectation we 
study numerically the effect of additional microwave
driving with dimensional amplitude $\epsilon_1$ ($\epsilon_1 \ll \epsilon$)
and frequency $\omega_1$. We use the wall model (W2) based on the
Chirikov standard map described here and in \cite{adcdls}.
As for the additional driving frequency we have 
$\epsilon_1=|E_1|/(\omega_1 v_F)$ where $E_1$ is
the field strength of microwave frequency $\omega_1$, we assume that both 
main and additional fields are collinear and perpendicular to the wall.
In presence of second frequency the map (\ref{eq3}) takes the form:
\begin{eqnarray}
\label{eq7}
{\bar v_y}  & = & v_y + 2 \epsilon \sin \phi  + 
  2 \epsilon_1 \sin(\omega_1 \phi /\omega ) +I_{cc}, \;\\
\nonumber
{\bar \phi} & = & \phi + 2 (\pi - {\bar v_y}) \omega_ /\omega_c \; .
\end{eqnarray}
The only modification appears in the first equation since now the 
change of velocity at collision depends on both fields;
the second equation remains the same as in (\ref{eq3}). 
As in the model (W2) the term $I_{cc}$ describes 
the effects of dissipation with rate $\gamma_c$ and noise with
amplitude $\alpha$ of random velocity angle rotations.  
We call this system model (W3).

\begin{figure}[!ht] 
\begin{center} 
\includegraphics[width=0.43\columnwidth]{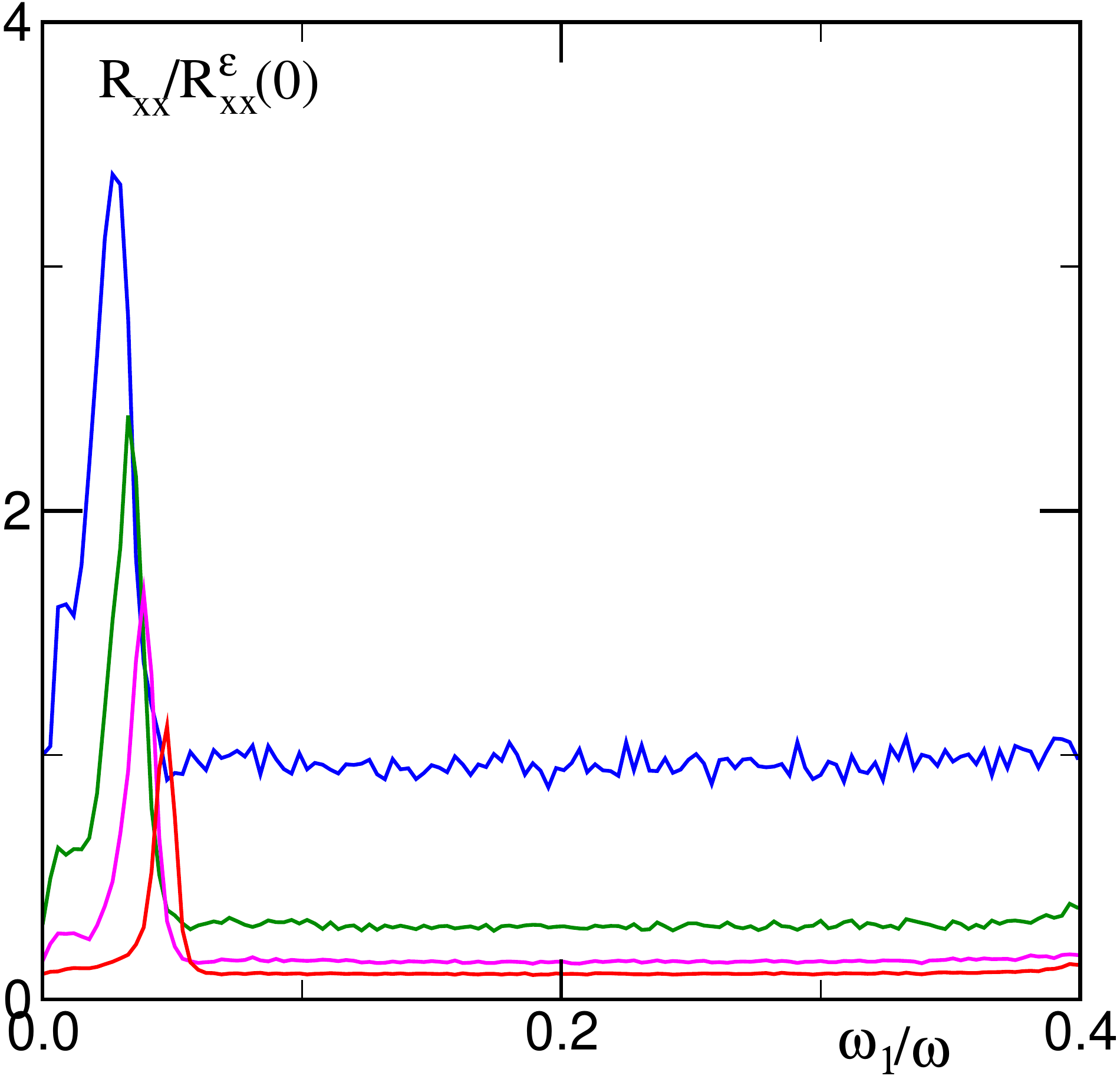}
\includegraphics[width=0.46\columnwidth]{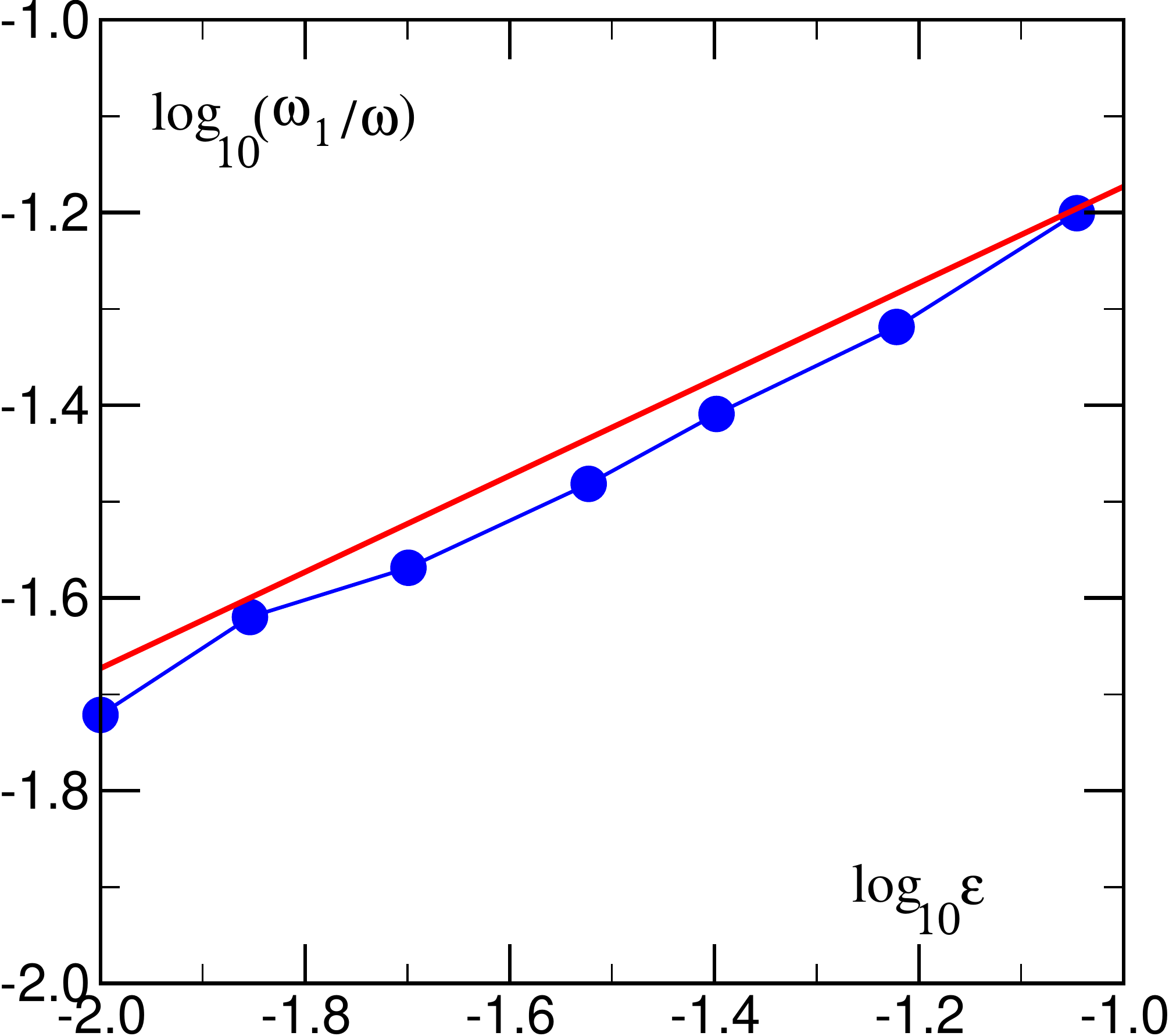}
\end{center}
\caption{(Color online) {\it Left panel:}  Dependence of
rescaled resistance $R _{xx}/R^\epsilon_{xx}(0)$
on frequency ratio $\omega_1/\omega$ in model (W3) 
described by the map (\ref{eq7}). 
Here, the test driving  at frequency $\omega_1$ has fixed amplitude 
$\epsilon_1=0.007$;
the main microwave driving 
is located in the ZRS phase at $j=\omega/\omega_c=2.25$,
and its amplitude takes values
$\epsilon = 0.02$ (blue curve), $0.03$ (green curve),
$0.04$ (magenta curve), $0.06$ (red curve) 
(these curves follow from top to bottom at $\omega_1/\omega=0.2$).
The values of $R_{xx}$ are computed at fixed $\epsilon_1=0.007$
and corresponding $\epsilon$; the values of 
$R^\epsilon_{xx}(0)$ are computed at $\epsilon_1=0$ and $\epsilon=0.02$.
The data are obtained at noise amplitude $\alpha=0.02$ and
dissipation $\gamma_c=0.01$; averaging is done over 2000
trajectories for 5000 iterations of map (\ref{eq7}).
{\it Right panel:} Dependence of peak position
$\omega_1/\omega$ on main microwave driving amplitude
$\epsilon$ obtained from data of left panel at $\epsilon_1=0.007$
and additional data at $\epsilon_1=0.003$ (blue points),
the theory dependence (\ref{eq6}) at $\omega/\omega_c=2.25$
is shown by the straight red line. Data for model (W3).
} 
\label{fig17}
\end{figure}

In the model (W3) the resistance  $R_{xx}$
is computed numerically in the same way as
in the model (W2) described in \cite{adcdls}:
the displacement along the edge between collisions
is $\delta x=2v_y/\omega_c$; it determines the total displacement
$\Delta x$  along the edge during the total computation
time $\Delta t \sim 10^4/\omega$; then 
$R_{xx} \propto 1/D_x = \Delta t/(\Delta x)^2$
where $D_x$ is an effective diffusion rate along the edge. 
To see the effect of additional weak test driving $\epsilon_1$
at frequency $\omega_1$ we place the system
in the ZRS phase at $j=\omega/\omega_c=2.25$ and measure the
variation of rescaled resistance
$R _{xx}/R^\epsilon_{xx}(0)$. Here $R_{xx}$
is the resistance in presence of 
both microwave fields $\epsilon$ and $\epsilon_1$
while $R^\epsilon_{xx}(0)$ is the resistance at
$\epsilon_1=0$ and a certain fixed $\epsilon$.
The dependence of $R _{xx}/R^\epsilon_{xx}(0)$
on the frequency ratio is shown in Fig.~\ref{fig17} at
left panel. The main feature of this data is appearance
of peak at low frequency ratio $\omega_1/\omega < 0.1$.
In the range $0.1 < \omega_1/\omega <0.4$ the testing field
$\epsilon_1$ is nonresonant and does not affect $R_{xx}$
however at $\omega_1/\omega < 0.1$ it becomes resonant to the 
pendulum oscillations in the wall vicinity and hence
strongly modifies $R_{xx}$ value. The dependence of 
this resonance ratio $\omega_1/\omega$ on 
amplitude of main driving field $\epsilon$ is shown in right panel
of Fig.~\ref{fig17}. The numerical data are in a good
agreement with the above theoretical expression (\ref{eq6}).

The theoretical dependence (\ref{eq6}) allows to check the 
synchronization theory of edge state stabilization.
It also allows to measure the strength of main microwave driving
force acting on an electron that still remains an experimental challenge. 
The experimental testing of relation (\ref{eq6}) 
requires to work with good ZRS samples which
have every low resistance in ZRS minima since this makes 
the effect of testing field $\omega_1$ to be more visible.
We note that the recent experiments in a low frequency regime
$\omega/\omega_c \ll 1$ \cite{alikcam} demonstrate that
$R_{xx}$ is sensitive to low frequency driving.
The expression (\ref{eq6}) is written for the case 
when $R_{xx}$ is mainly determined by a transport along
edges. If the dominant contribution is given
by bulk disk scatterers then a certain numerical
coefficient $A$ should be introduced in the right part
of the expression. According to the data of Fig.~\ref{fig8}
and Eqs.~(\ref{eq5})
we estimate $A \approx 0.5$ (the separatrix width is
smaller for the disk case compared to the wall case at the same
$\epsilon$).   

Another interesting experimental possibility of our theory
verification is to take a Hall bar of high mobility 2DEG sample
and put on it antidots with regular or disordered
distribution (it is important to have no direct
collision-less path for a cyclotron radius 
in crossed {\it dc-}electric and magnetic 
fields) with a low density of antidot disks
$n_d r_d^2 \ll 1$ (as in our numerical studies)
so that an average distance between antidots is 
larger than the cyclotron radius $r_c$.
The regular antidot lattices have been already realized experimentally
\cite{weiss,kvon}. The effect of microwave field on electron transport
in a regular lattice has been studied in the frame of
ratchet transport in asymmetric lattices \cite{portal}.
Even a case of  symmetric circular antidots  has been studied in 
\cite{portal} but the lattice was regular and no special 
attention was paid to
analysis of resistivity at ZRS resonant regime with $j \approx j_r$.
We think that the experimental condition 
of  \cite{portal} can be relatively easy modified
to observe the ZRS effect on disk scatterers discussed here.

\section{Discussion}

Above we presented theoretical and numerical results
which in our opinion explain the appearance of 
microwave induced ZRS in high mobility samples.
The synchronization theory of ZRS proposed in \cite{adcdls}
and extended here is based on a clear physical picture:
high harmonics $\omega_/\omega_c=j > 1$ 
are generated by collisions with sharp
edge boundary or isolated impurities which are modeled here by
specular disks. The ZRS phases appear in a vicinity of 
resonant values $j_r \approx 1+1/4, \; 2+1/4, \; ...$ .
At these $j_r$ values the cyclotron phase of electron motion
becomes synchronized with the microwave phase
due to dissipative processes present in the system.

For trajectories at the edge vicinity this synchronization
gives stabilization of propagation along edge channels
that creates an exponential drop of resistivity 
contribution of these channels with decreasing amplitude of
thermal noise and increasing amplitude of microwave field.
The contribution to resistivity from trajectories
in the bulk is analyzed in the frame of 
scattering on many well separated disk impurities.
Here again the synchronization of cyclotron 
phase with the microwave phase takes place approximately
at the same resonant $j_r$ values. At these $j_r$ values
the synchronization leads to long time capture of trajectories
in disk vicinity. During this long time an initial 
cyclotron phase is washed out by noisy fluctuations
and many rotations around disk and thus
an electron escapes from a disk with an average zero
displacement along the applied {\it dc-}field
even if dynamics in disk vicinity is dissipative. 
This provides the main mechanism of
suppression of dissipative resistivity contribution from
isolated impurities in the bulk. 
As a result the contribution of bulk to 
dissipative conductivity $\sigma_{xx}$
is suppressed, as it was assumed in \cite{adcdls},
and the main contribution to  current is given by
electron propagation along edge states 
stabilized by a microwave field.

As we showed above the resonance width
or resonance energy scale $E_r$ are approximately 
the same for the disk  and wall cases
(see Eqs.~(\ref{eq4},~\ref{eq5})).
We note that for the disk case the energy $E_r$
is not sensitive to the disk radius as soon as
it is significantly smaller than the cyclotron radius.
Thus we expect that
at $j_r$ values the conductivity $\sigma_{xx}$ in the bulk
is suppressed by a microwave field
and at  these fields
the current is flowing essentially along 
stabilized edge states. In the case of
Corbino geometry we have 
radial conductivity  $\sigma_{rr}$
which is determined by the bulk
scattering and now the minima of $\sigma_{rr}$
are located at $j_r$ values (see e.g Figs.~\ref{fig12},
\ref{fig13},\ref{fig14} where $R_{xx} \propto \sigma_{xx} \sim \sigma_{rr}$).
The ZRS experiments performed in the  Corbino geometry
give minima of $\sigma_{rr}$ at these $j_r$ values
(see e.g. \cite{zudovcorbino,bykovcorbino}) being
in agreement with the synchronization theory.

It is interesting to note that
the nonlinear dynamics in vicinity of edge and disk impurity
is well described by the Chirikov standard map \cite{chirikov}.
The map description explains the location of resonances
at integer values of $j$ with an additional shift $\delta j \approx 1/4$
produced by a finite separatrix width of nonlinear resonance.
A finite rigidity of wall or disk potential can give
a modification of this shift $\delta j$.

Our results show that the ZRS phases at $j_r$ appear only at weak noise
corresponding to high mobility samples.
Strong noise destroys synchronization and trajectories are no more
captured at edge or disk vicinity. We also note that
internal sample potentials with significant gradients
act like a strong local {\it dc-}field which destroys
stability regions around disk impurities or near edge.
Due to that the ZRS effect exists only in high mobility samples.
The resistance at ZRS minima drops significantly 
with the growth of microwave field strength since
it increases the amplitude of nonlinear resonance 
which captures the synchronized trajectories.

The synchronization theory of ZRS is based on classical dynamics 
of noninteracting electrons. It is possible that electron-electron interaction 
effects can also suppress the contribution to resistivity 
from neutral short range range scatterers (interface roughness, adatoms,...). 
Indeed, long capture times can increase the electron density
around these short ranged impurities transforming them 
into long range charged scatters that the other electrons 
can circumvent by adiabatically following the long range component 
of the disorder potential thereby avoiding a scattering event.
However, the theoretical description of this short-ranged impurity cloaking 
mechanism for the ZRS effect remains a serious challenge.

Another important step remains the development of
a quantum synchronization theory for ZRS. 
Even if in experiments the Landau level
is relatively high $\nu \sim 60$, there are
only about ten Landau levels inside a nonlinear resonance \cite{adcdls}
and  quantum effects should play a significant role.
The general theoretical studies show that
the phenomenon of quantum synchronization 
persists at small effective values of Planck constant $\hbar_{eff}$
but it becomes  destroyed by 
quantum fluctuations at certain large values of $\hbar_{eff}$ \cite{qsync}.

The importance of quantum ZRS theory is also related to 
a short range nature of the impurities considered here, 
typically on a scale of a few nanometers or even less. 
We have modeled these 
impurities by disks with a radius that was 
only several times (in fact $j$ times) smaller
than the cyclotron radius which is not so close to 
microscopic reality. We could argue that
in the quantum case a nanometer sized impurity
would act effectively as an impurity of a size
of quantum magnetic length 
$a_B \sim r_c/\sqrt{\nu} \approx r_c/8 \sim 100 nm$.
This gives a ratio $r_c/a_B \sim 8$ 
which is comparable with the one used in our simulations
with  $ r_c/r_d = j \sim 7$ but of course a 
quantum treatment of scattering on nanometer size impurity
in crossed electric and magnetic and also microwave fields
remains a theoretical challenge. We note that
such type of scattering can be efficiently
analyzed by tools of quantum chaotic scattering
(see e.g. \cite{martina,main}) and we expect that these
tools will allow to make a progress in
the quantum theory development of striking ZRS phenomenon.

We hope that the synchronization theory of microwave induced ZRS
phenomenon described here can be tested in further ZRS experiments.

\section{Acknowledgments}
This work was supported in part by ANR France PNANO project NANOTERRA;
OVZ was partially supported by the Ministry of
Education and Science of  Russian Federation.

We dedicate this work  to the memory of Boris Chirikov
(06.06.1928 - 12.02.2008).

%\newpage

\end{document}